\documentclass{pasj02}
\usepackage{url}
\usepackage{diagbox}

\Received{$\langle$reception date$\rangle$}
\Accepted{$\langle$acception date$\rangle$}
\Published{$\langle$publication date$\rangle$}

\begin{document}

\title{Thermal evolution model from cometary nuclei to asteroids considering contraction associated with ice sublimation}

\author{Hitoshi Miura$^{1*}$ and Takumi Yasuda$^2$}% \orcid{0000-0001-8891-1658}}%

\altaffiltext{1}{Graduate School of Science, Nagoya City University, Yamanohata 1, Mizuho-cho, Mizuho-ku, Nagoya, Aichi 467-8501, Japan}
\altaffiltext{2}{MHI Aerospace Systems Corp., 10 Oe-cho, Minato-Ku, Nagoya, Aichi 455-8515, Japan}
\email{miurah@nsc.nagoya-cu.ac.jp}

\KeyWords{comets: general --- minor planets, asteroids: general --- meteorites, meteors, meteoroids }

\maketitle

\begin{abstract}
   Comet--asteroid transition (CAT) objects are small solar system bodies in the process of evolving from cometary nuclei into asteroids, as they gradually lose volatile substances due to solar heating. The volatile material is mainly water ice, and the time required for its complete depletion is called the desiccation time. Estimating the desiccation time is important for examining the formation and evolution of small solar system bodies. Here, we propose a new theoretical model for evaluating the desiccation time as a function of orbital elements, considering the contraction of the entire cometary nucleus due to ice sublimation. First, we performed numerical calculations of the thermal evolution of a cometary nucleus in an eccentric orbit, considering the seasonal variation in the solar heating rate. Next, we derived the desiccation time analytically as a function of orbital elements based on a steady-state model considering the solar heating rate averaged over the seasons. We compared the numerical solutions for the desiccation time with the analytical solutions and clarified the conditions under which the analytical model can be applied. Additionally, based on the analytical model, we derived formulae for estimating the emission rates of water vapor and dust on the surface of the cometary nucleus, the maximum size of the emitted dust, and the dust emission velocity, by assuming the amount of ice remaining inside the nucleus. Using these analytical solutions, we considered the internal structure and evolution process of typical CAT objects. Our analytical model was generally consistent with that of the results of earlier observations of these objects. Our model provides a theoretical guideline for discussing the evolution of cometary nuclei and the possibility of retaining internal ice in asteroids.
\end{abstract}

% \pagewiselinenumbers

\section{Introduction}

In addition to the planets, dwarf planets and their satellites, various other small celestial bodies exist in the solar system. These small celestial bodies are mainly classified according to their orbital elements, and are roughly divided into two categories: asteroids, which are found within the region around the orbit of Jupiter, and trans-Neptunian objects (TNOs), which are found beyond the orbit of Neptune. Comets are also included in the category of TNOs. Asteroids, comets and other small celestial bodies have not undergone much of internal alteration processes, and they may retain information from the early stages of the formation of the solar system. They are called primordial celestial bodies, and have been the target of exploration aimed at obtaining clues to help elucidate the history of the solar system.

In the classical classification, comets were thought to be the celestial bodies rich in volatile materials originating from outside the orbit of Neptune, and showed characteristic phenomena such as comas and tails (cometary activity) due to the sublimation of the composed volatiles. Contrary, asteroids were considered to be inactive celestial bodies composed of non-volatile materials forming within the orbit of Jupiter. However, with the improvement of observational instruments, celestial bodies with characteristics of both comets and asteroids were discovered, and the distinction is considered to be very vague \citep{Hsieh2017}. One of the factors that has led to the blurring of the distinction between the two is the evolution of comets into asteroids due to solar heating. When a comet approaches the Sun and is heated, the volatile substances (hereafter referred to as ``ice'') inside the cometary nucleus sublimate and are released into space. Some of the non-volatile substances (hereafter referred to as ``rock'' or ``dust'') contained in the cometary nucleus were blown away from the nucleus due to the outflowing gas, resulting in the formation of coma and dust tails. However, rocks of a certain size are attracted by the gravity of the cometary nucleus and remain on the surface. As a result, a layer of rocks (dust mantle) forms on the inner region where ice remains (primitive core). Because the dust mantle is porous, the gas forming in the primitive core permeates the dust mantle and flows out into space. Eventually, the ice inside the nucleus is lost, and it evolves into an asteroid made up mainly of rock. Small bodies in the middle of this evolutionary stage are called comet--asteroid transition (CAT) objects \citep{Hsieh2004AJ}.

The evolution of cometary nuclei due to solar heating was investigated using a number of theoretical models. The typical models are summarized in Table \ref{table:models}. The models before 2000 AD mainly assumed short-period comets (SPCs). As SPCs follow highly eccentric orbits, solar heating is only significant near perihelion. \citet{Mendis1977aa} calculated the time variation of the dust mantle thickness due to ice sublimation, and showed that after perihelion, the ice sublimation is suppressed by the insulating effect of the thick dust mantle. \citet{Brin1979ApJ} showed that the dust mantle is catastrophically destroyed before perihelion passage because large rocks of several centimeters are blown away by the outflow gas. \citet{PANALE1984476} examined the effect of the dust mantle on the suppression of water vapor outflow, and showed that a dust mantle with a thickness of several centimeters would not be blown away (permanent mantle formation), and would irreversibly evolve into an asteroid. Later, models were proposed considering not only the porocity of the dust mantle but also of the primitive core, and the effects of water vapor transport into the interior of the cometary nucleus were examined \citep{Mekler1990ApJ,Prialnik1990ApJ,ESPINASSE1991350}. In the 2000s, models taking into account the relatively small orbital eccentricity of asteroidal orbits were considered, as small bodies with orbital orbits like asteroids but exhibiting comet-like activity were discovered (main belt comets, MBCs). The difference from SPCs is that MBCs are continuously heated at all positions on their orbital path, not just at perihelion. \citet{Prialnik2009MNRAS} performed long-term evolution calculations over a period of 4.6 billion years, assuming comet 133P/Elst-Pizarro, the first MBCs discovered, and showed the possibility that crystalline ice remains inside the nucleus. \citet{Schorghofer2018JGR} and \citet{Yu2019MNRAS} analytically derived the time required for the depletion of internal ice (desiccation time), and formulated the conditions for retaining internal water ice as a function of orbital elements. \citet{Miura2022ApJ} examined a theoretical model considering the contraction of the cometary nucleus due to ice sublimation, and showed that asteroid Ryugu with a radius of about 440 m may have evolved from a cometary nucleus with a radius of 1.2 km over a period of about tens of thousands of years. The theoretical model described above assumed a fixed orbital path, although reports of studies exist considering the time evolution of the orbital path \citep{CORADINI1997337,SCHORGHOFER2020113865}.

Estimating the desiccation time of cometary nuclei is important for discussing the possibility that the small bodies under consideration are of cometary origin or that they may still contain ice inside. \citet{Schorghofer2018JGR} and \citet{Yu2019MNRAS} showed that the desiccation time can be evaluated from the solar radiation averaged over the orbital period. However, their models did not consider the contraction of the cometary nucleus associated with ice sublimation. When ice sublimates, the rocks trapped within the ice lose their support and fall, causing the entire cometary nucleus to contract. The contraction may change the porosity of the dust mantle, and thus the thermal conductivity and permiability, which may affect the estimation of the desiccation time. \citet{Miura2022ApJ} derived an analytical solution for the desiccation time based on a model considering the contraction of the cometary nucleus, however, they did not consider the thermal evolution. Furthermore, although the theoretical models mentioned above all assumed that the temperature of the primitive core is constant and uniform during the evolution process, this is not self-evident.

The study objective is to derive the desiccation time by considering the contraction of the cometary nucleus. First, we describe a numerical model of the thermal evolution of cometary nuclei based on the theoretical model of \citet{Miura2022ApJ} (see section \ref{sec:numerical_model}). In the numerical model, we consider the seasonal cycle in the solar heating rate. Next, assuming that the temperature and distribution of water vapor inside the cometary nucleus reached a steady state, we derived an analytical solution for the desiccation time (see section \ref{sec:steady_model}). In the analytical model, we used the seasonal average of the solar heating rate. We performed numerical calculations of the thermal evolution of cometary nuclei for various orbital elements, and clarified the conditions for the applicability of the analytical model by comparing the numerical solutions for the desiccation time with the analytical solutions (see section \ref{sec:results}). We discussed the internal structure and the evolution of small solar system bodies by comparing our theory with that of the observational results for typical CAT objects (see section \ref{sec:discussion}). Finally, the conclusions are summarized in section \ref{sec:conclusion}.

\begin{table*}
   \footnotesize
   \caption{Theoretical models of the thermal evolution of cometary nuclei due to solar heating and the physical processes considered in each. The physical processes considered are indicated by ``X". The meanings of each process are as follows. \textbf{Formation}: formation of dust mantle, \textbf{Blown-off}: the blowing away of dust mantle by outflowing gas, \textbf{Insulation}: insulating effect of the dust mantle, \textbf{Cap}: effect of the dust mantle in suppressing gas outflow, \textbf{Collapse}: contraction of the cometary nucleus due to sublimation of ice, \textbf{Permeability}: vapor permeation flow within the primitive core, \textbf{Multi}: sublimation of multi-component ices (other than H$_2$O), \textbf{Amor}: crystallization of amorphous H$_2$O ice. \textbf{Orbit} indicates the type of orbital elements considered. \textbf{SPC}: short-period comet, \textbf{MBC}: main-belt comet, \textbf{NEA}: near-earth asteroid, \textbf{Change}: time evolution of orbital elements is considered, \textbf{Free}: orbital elements are free parameters.}
   \begin{center}
      \begin{tabular}{l|ccccc|c|cc|c}
         \hline
         & \multicolumn{5}{c|}{Dust mantle} & Primitive core & \multicolumn{2}{c|}{Ice model} & \\
         \hline
         Literature & Formation & Blown-off & Insulation & Cap & Collapse & Permeability & Multi & Amor & Orbit \\
         \hline 
         \citet{Mendis1977aa} & X & & X & & & & & & SPC \\
         \citet{Brin1979ApJ} & X & X & X & & & & & & SPC \\
         \citet{PANALE1984476} & X & X & X & X & & & & & SPC \\
         \citet{Mekler1990ApJ} & & & & & & X & & & SPC \\
         \citet{Prialnik1990ApJ} & & & & & & X & X & X & SPC \\
         \citet{ESPINASSE1991350} & & & & & & X & X & X & SPC \\
         \citet{CORADINI1997337} & X & X & & & & X & X & X & Change \\
         \citet{Prialnik2009MNRAS} & X & X & X & X & & & X & X & MBC \\
         \citet{Schorghofer2018JGR} & X & & X & X & & & & & Free \\
         \citet{Yu2019MNRAS} & X & & X & X & & & & & Free \\
         \citet{SCHORGHOFER2020113865} & X & & X & X & & & & & Change \\
         \citet{Miura2022ApJ} & X & & & X & X & X & & & NEA \\
         \hline
         This study & X & & X & X & X & X & & & Free \\
         \hline
      \end{tabular}\label{table:models}
   \end{center}
\end{table*}

\section{Numerical model}

\label{sec:numerical_model}

\subsection{Outline}

\begin{figure*}
   \begin{center}
      \includegraphics[width=\linewidth]{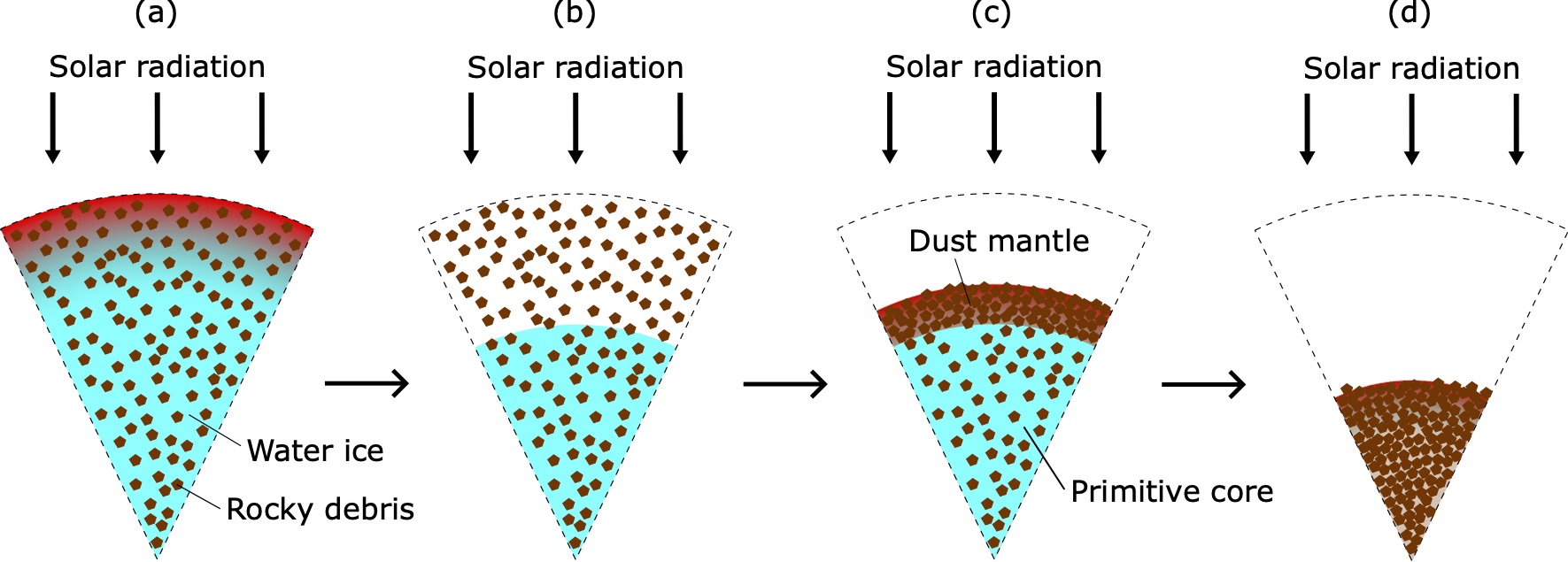}
   \end{center}
   \caption{Numerical model of the thermal evolution from cometary nuclei to asteroids heated by sunlight. (a) Internal structure of a cometary nucleus before sublimation of ice. It is composed of water ice and rock. The surface is heated by solar radiation, and heat is transmitted to the interior by solid heat conduction. (b) Ice sublimates, and the rock buried in the ice is exposed. (c) The exposed rock falls under the gravitational force of the nucleus itself, and due to the simultaneous formation of a porous dust mantle on its surface, the entire nucleus contracts. The water vapor generated by the primitive core flows outwards through the pores in the dust mantle. The primitive core itself is also porous, and the movement of water vapor towards the center is also considered. (d) The ice inside is depleted, and the object comprises only rock, namely, asteroid.
   {Alt text: Illustrations.}}
   \label{fig:Model}
\end{figure*}

The numerical model of the thermal evolution of a cometary nucleus is shown in Figure \ref{fig:Model}. Here, only water ice was considered as a volatile substance. The initial state is a spherical cometary nucleus with a porous internal structure randomly packed with ice particles and rocky debris (a). The ice particles and rocky debris are sufficiently mixed, and the interior of the cometary nucleus is considered to be uniform on a macroscopic scale. When the cometary nucleus is heated from the surface by sunlight, the temperature near the surface rises and the ice sublimes, filling the pores inside the nucleus with water vapor. The pores are connected, and the water vapor moves between the pores according to the gradient of the water vapor pressure distribution. Near the surface of the cometary nucleus, the water vapor is transported outward and released into space. Therefore, the water vapor pressure inside the nucleus decreases, and the internal ice continuously sublimes. When the ice sublimes, the rocks trapped within it are exposed (b). The rocks, which have lost their support from the ice, accumulate on the surface of the nucleus to form a porous dust mantle (c). Here, only rocks with a diameter of 1 m were considered, and it was assumed that all the rocks would accumulate on the surface of the nucleus. Additionally, the entire cometary nucleus contracts as the rocks fall. The water vapor produced by the internal ice permeates the porous dust mantle and flows out into space, which finally leaves, only a small celestial body made of rocks (d).

We assumed that the structure of the cometary nucleus is spherically symmetric, and that the spatial distribution of physical quantities depends only on the distance $r$ from the center. We assumed that the sunlight energy received by the nucleus is averaged over the entire surface, and ignored any differences due to latitude or longitude. Assuming that the cometary nucleus revolves in a fixed eccentric orbit with the Sun as one of its focal points, we considered the seasonal cycle of the solar flux corresponding to the change in the distance between the Sun and the nucleus.

\subsection{Time evolution of internal structure}
\label{sec:basic_equation}

Let $T(r)$ be the temperature inside the cometary nucleus, $\rho_\mathrm{v}(r)$ be the density of water vapor, and $\phi_\mathrm{i}(r)$ and $\phi_\mathrm{r}(r)$ be the volume fractions occupied by ice and rock, respectively. The subscripts v, i, and r indicate that they are physical quantities for vapor, ice, and rock, respectively. The porosity is given by $\epsilon = 1 - \phi_\mathrm{i} - \phi_\mathrm{r}$. The density of vapor, $\rho_\mathrm{v}$, does not represent the density in the pore, but rather the mass of vapor per unit volume of the cometary nucleus. The pressure $P$ of the water vapor filling the pores is related to the ideal gas equation of state: $\rho_\mathrm{v} = \epsilon m_\mathrm{v} P/(k_\mathrm{B}T)$, where, $m_\mathrm{v}$ and $k_\mathrm{B}$ are the mass of a water molecule and Boltzmann's constant, respectively. The difference in temperature between the components inside the nucleus is ignored (local thermal equilibrium). The time variations of $T(r)$ and $\rho_\mathrm{v}(r)$ are obtained by solving the following spherically-symetric heat conduction equation and mass conservation equation, respectively:
\begin{equation}
    \rho c \frac{\partial T}{\partial t}
    = \frac{1}{r^2} \frac{\partial}{\partial r} \left( r^2 \kappa \frac{\partial T}{\partial r} \right) - q_\mathrm{v} H_\mathrm{i} ,
    \label{eq:energy_conservation01}
\end{equation}
\begin{equation}
    \frac{\partial \rho_\mathrm{v}}{\partial t}
    = - \frac{1}{r^2} \frac{\partial}{\partial r} \left(r^2 J_\mathrm{v} \right) + q_\mathrm{v} .
    \label{eq:vapor_conservation01}
\end{equation}
Here, $\rho$, $c$, and $\kappa$ are the density, specific heat, and thermal conductivity of the cometary nucleus, respectively, $q_\mathrm{v}$ is the source term of water vapor, $H_\mathrm{i}$ is the specific sublimation heat of ice, and $J_\mathrm{v}$ is the mass flux of water vapor passing through a unit cross-sectional area of the cometary nucleus. It was assumed that only the solid components, ice and rock, contributed to the heat capacity, and it was set to be $\rho c = \varrho_\mathrm{i} c_\mathrm{i} \phi_\mathrm{i} + \varrho_\mathrm{r} c_\mathrm{r} \phi_\mathrm{r}$, where, $\varrho$ is the material density of the solid component. The value of $\kappa$ was given based on Russel's formula, which considers the solid thermal conductivity and radiative heat transfer within porous materials \citep{Orosei1995AA}. We ignored the radiative heat transfer in our model and set the value of $\kappa$ to be equal to the solid thermal conductivity, i.e., $\kappa = (1 - \epsilon^{2/3})/(1 - \epsilon^{2/3} + \epsilon) \bar{\kappa}_\mathrm{s}$. Here, $\bar{\kappa}_\mathrm{s}$ is the thermal conductivity of the solid, and is given by taking a weighted average of the thermal conductivities of ice and rock by volume fraction, e.g., $\bar{\kappa}_\mathrm{s} = (\phi_\mathrm{i} \kappa_\mathrm{i} + \phi_\mathrm{r} \kappa_\mathrm{r})/(\phi_\mathrm{i} + \phi_\mathrm{r})$. According to the Hertz-Knudsen equation, $q_\mathrm{v}$ is given by
\begin{equation}
    q_\mathrm{v} = 
    \frac{6 \phi_\mathrm{i}}{d_\mathrm{i}}
    \left( \frac{m_\mathrm{v}}{2 \pi k_\mathrm{B} T} \right)^{1/2}
    \big[ P_\mathrm{i} (T)-P \big],
\end{equation}
where, $d$ and $P_\mathrm{i}(T)$ are the diameter of the solid component and equilibrium vapor pressure of water ice, respectively. $J_\mathrm{v}$ is given by \citep{Mekler1990ApJ}
\begin{equation}
    J_\mathrm{v} = - D_\mathrm{v}
    \frac{\partial}{\partial r} \left( \frac{P}{\sqrt{T}} \right) ,
    \label{eq:vapor_flux01}
\end{equation}
where, $D_\mathrm{v} = (16/3) (m_\mathrm{v}/(2 \pi k_\mathrm{B}))^{1/2} \epsilon^{3/2}/(1 - \epsilon)^{1/3} \bar{d}_\mathrm{s}$ represents the diffusion efficiency of water vapor in a porous medium. The average grain diameter of the solid components, $\bar{d}_\mathrm{s}$, was given by using a weighted average based on the volume fraction, as follows: $\bar{d}_\mathrm{s} = (\phi_\mathrm{i} d_\mathrm{i} + \phi_\mathrm{r} d_\mathrm{r})/(\phi_\mathrm{i} + \phi_\mathrm{r})$.

The time variation of $\phi_\mathrm{i}$ due to ice sublimation is given by
\begin{equation}
    \varrho_\mathrm{i} \frac{d \phi_\mathrm{i}}{dt} = - q_\mathrm{v} .
    \label{eq:sublimation01}
\end{equation}
When $P > P_\mathrm{i}$, $q_\mathrm{v} < 0$, which implies that the volume of ice increases as the water vapor condenses. The condensation of water vapor acts as a positive heat source (condensation heat) in the heat conduction equation (\ref{eq:energy_conservation01}).

When the ice sublimates, the rocks buried within it are exposed. The rocks, which have lost their support from the ice, fall towards the center of the cometary nucleus, pulled by the gravity. Therefore, the entire cometary nucleus contracts. The extent of contraction should be determined by considering the balance between the self-gravity and internal strength of the porous celestial body. For simplicity, our model sets a maximum porosity value, $\epsilon_\mathrm{max}$, and treats the case where the porosity exceeds this value, i.e., the porosity in that region is compressed until it satisfies the condition of $\epsilon = \epsilon_\mathrm{max}$ \citep{Miura2022ApJ}. We took the typical porosities of cometary nuclei and asteroids to be $\epsilon_\mathrm{p}$ and $\epsilon_\mathrm{m}$, respectively, and assigned $\epsilon_\mathrm{max}$ the value $\epsilon_\mathrm{max} = (\phi_\mathrm{i} \epsilon'_\mathrm{p} + \phi_\mathrm{r} \epsilon_\mathrm{m})/(\phi_\mathrm{i} + \phi_\mathrm{r})$ by taking the weighted average of the volume fraction of ice and rock so that $\epsilon_\mathrm{max} = \epsilon_\mathrm{p}$ in the primitive region before ice sublimation and $\epsilon_\mathrm{max} = \epsilon_\mathrm{m}$ in the region of rock alone where ice was completely depleted. Here, we define $\epsilon'_\mathrm{p}$ so that it satisfies the equation: $(\phi_\mathrm{i0} \epsilon'_\mathrm{p} + \phi_\mathrm{r0} \epsilon_\mathrm{m})/(\phi_\mathrm{i0} + \phi_\mathrm{r0}) = \epsilon_\mathrm{p}$, where, $\phi_\mathrm{i0}$ and $\phi_\mathrm{r0}$ are the initial volume fractions of ice and rock before ice sublimation, respectively. The specific numerical calculation method for the contraction of the cometary nucleus is explained in section \ref{sec:numerical}.

\subsection{Heat flux on eccentric orbit}

On the surface of a cometary nucleus, heating due to sunlight and cooling due to the thermal radiation exists. If the distance between the nucleus and the Sun is $l$ (au), the amount of solar energy received per unit time and unit surface area is given by $\Phi_\odot(l) = (1-A) S_\odot / (4 l^2)$. Here, $A$ is the albedo of the nucleus surface, and $S_\odot$ is the solar constant. The energy received by the cross-sectional area of $\pi R_*^2$ is averaged over the entire surface area of $4 \pi R_*^2$, where, $R_*$ is the entire nucleus radius. Contrary, cooling due to thermal radiation is given by the Stefan-Boltzmann law as follows: $\Phi_\mathrm{rad} = \sigma_\mathrm{SB} T^4(R_*)$. Here, $\sigma_\mathrm{SB}$ is the Stefan-Boltzmann constant. Therefore, the net energy flux recieved by the nucleus is given by
\begin{equation}
    \Phi(l) = \Phi_\odot(l) - \Phi_\mathrm{rad}
    = \sigma_\mathrm{SB} \big[ T^4_\mathrm{eq}(l) - T^4(R_*) \big] ,
    \label{eq:energy_input01}
\end{equation}
where, $T_\mathrm{eq}(l)$ represents the surface temperature when heating and cooling are in balance (radiation equilibrium temperature), and is given by the following from the condition of $\Phi_\odot = \Phi_\mathrm{rad}$:
\begin{equation}
    T_\mathrm{eq}^4(l) = \frac{(1-A) S_\odot}{4 \sigma_\mathrm{SB} l^2} .
    \label{eq:T_eq01}
\end{equation}

As the cometary nucleus traces an eccentric orbit with the Sun as one of its focal points, $l$ changes over time. If the orbital semi-major axis is $a$ (au) and the orbital eccentricity is $e$, $l$ is given by the following as a function of the eccentric anomaly angle, $\theta$:
\begin{equation}
    l = a (1-e \cos \theta).
    \label{eq:heliocentric01}
\end{equation}
If the cometary nucleus is located at the aphelion ($\theta = \pi$) at time $t = 0$, the relationship between $t$ and $\theta$ satisfies the following Kepler equation:
\begin{eqnarray}
    \theta - e \sin \theta = \frac{2\pi}{\tau} t + \pi ,
    \label{eq:Kepler01}
\end{eqnarray}
where, $\tau$ is the orbital period.

\subsection{Numerical methods}
\label{sec:numerical}

\begin{figure}
   \begin{center}
      \includegraphics[width=60mm]{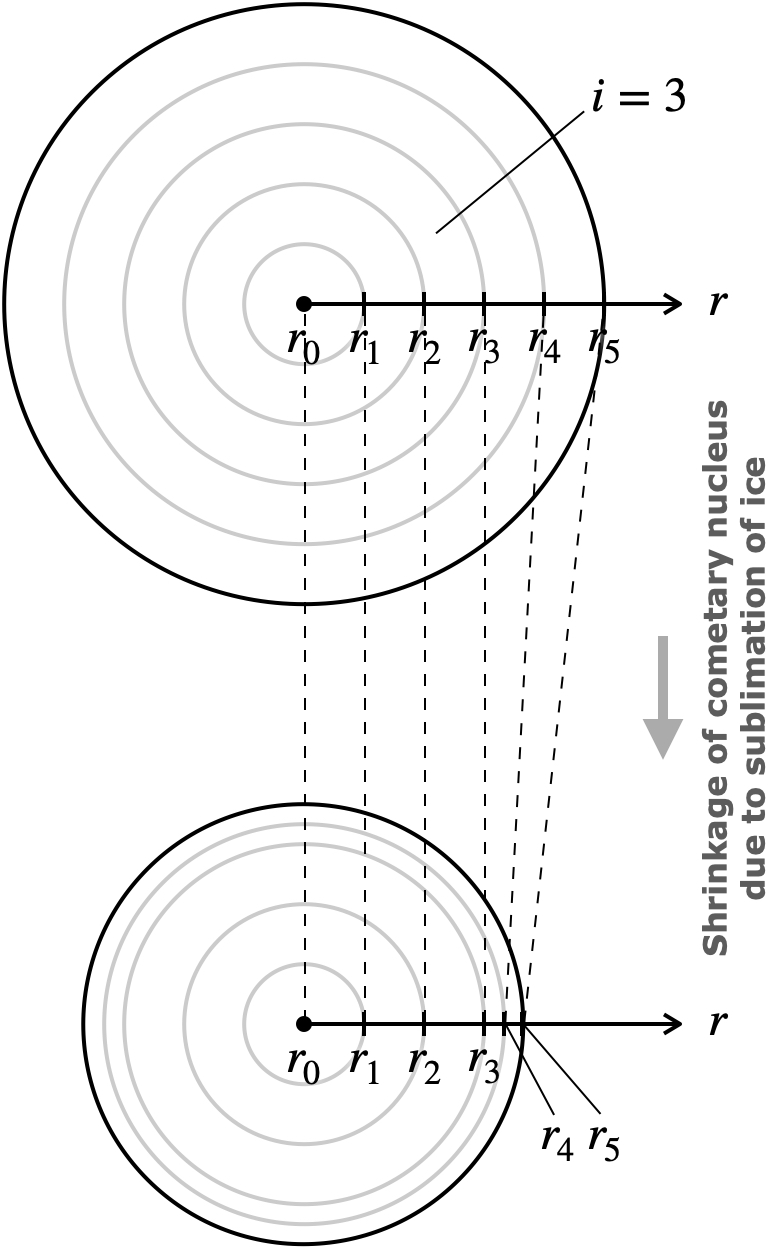}
   \end{center}
   \caption{Numerical method for descritizing the internal structure of a cometary nucleus and calculating its contraction due to ice sublimation. The interior of a cometary nucleus, which has a spherically symmetric structure, is divided into $N$ spherical shell cells with equal width at the beginning, and numbered from the inside out as $i=1,2,\cdots, N$ (in the figure, $N = 5$ is shown as an example). Only cell $i=1$ is a sphere rather than a spherical shell. The right boundary of cell $i$ is assigned the same number $i$ as the cell. The left boundary of the innermost cell $i=1$ is $i=0$. The distance from the center of the nucleus to the boundary $i$ is $r_i$. $r_0 = 0$ is the center of the nucleus, and $r_N = R_*$ is the radius of the nucleus. The contraction of the cometary nucleus is expressed by moving the position $r_i$ of each boundary.
   {Alt text: Illustrations.}}
   \label{fig:sabun}
\end{figure}

Figure \ref{fig:sabun} shows the finite difference method used for the interior of the cometary nucleus. The interior is divided into $N$ spherical shell cells, numbered $i = 1, 2, ..., N$ from the center outwards. The outer boundary of cell $i$ is denoted by $i$, and the distance from the nucleus center to boundary $i$ is denoted by $r_i$. The inner boundary of cell $i=1$ is $i=0$. $r_0$ denotes the position of the nucleus center and is always 0, and $r_{N}$ is equal to the radius $R_*$ of the cometary nucleus. At the beginning of the calculation, the thickness of cell $i$, $\Delta r_i = r_{i}-r_{i-1}$, is assumed to be equal for all cells. If the initial radius of the cometary nucleus is $R_0$, then the initial thickness of cell $i$ is given by $\Delta r_i = R_0/N$. After discretizing the interior accordingly, we applied the finite volume method to the spatial derivatives in equations (\ref{eq:energy_conservation01}) and (\ref{eq:vapor_conservation01}), and the first-order implicit backward difference method to the time derivatives. At $r = r_N$, we imposed the boundary conditions of net energy inflow $4 \pi r_N^2 \Phi(l)$ (see equation \ref{eq:energy_input01}) and $P = 0$. At $r = 0$, the gradients of $T$ and $P$ were both set to zero. The time variation of $\phi_\mathrm{i}$ was calculated using an analytical solution considering the fact that $q_\mathrm{v}$ is proportional to $\phi_\mathrm{i}$ in equation (\ref{eq:sublimation01}).

The contraction of the cometary nucleus due to ice sublimation is expressed by changing $r_i$. If the position of cell boundary $i$ before and after contraction are $r_i^{(n)}$ and $r_i^{(n+1)}$, respectively, the volume ratio of cell $i$ after contraction to that before, $\chi_i$ ($< 1$), is given by
\begin{equation}
    \chi_i = \frac{(r_{i}^{(n+1)})^3-(r_{i-1}^{(n+1)})^3}{(r_{i}^{(n)})^3-(r_{i-1}^{(n)})^3} .
    \label{eq:shrinkage01}
\end{equation}
The value of $\chi_i$ is determined so that the porosity of each cell is equal to $\epsilon_\mathrm{max}$ (see section \ref{sec:basic_equation}). First, suppose that the volume fraction of ice in cell $i$ decreases from $\phi_{\mathrm{i},i}$ to $\phi'_{\mathrm{i},i}$ due to sublimation. As we are ignoring sublimation of the rock, $\phi_{\mathrm{r},i}$ does not change at this time. Due to cell contraction the volume fractions of ice and rock become $\phi'_{\mathrm{i},i}/\chi_i$ and $\phi_{\mathrm{r},i}/\chi_i$, respectively. Therefore, the following is satisfied with respect to porosity after cell contraction:
\begin{equation}
    1 - \frac{\phi'_{\mathrm{i},i} + \phi_{\mathrm{r},i}}{\chi_i} = \epsilon_\mathrm{max} .
    \label{eq:shrinkage02}
\end{equation}
From equation (\ref{eq:shrinkage02}), we can determine the value of $\chi_i$ for all cells. Using the fact that $r_0^{(n+1)} = 0$, we can find $r_{1}^{(n+1)}, r_{2}^{(n+1)}, r_{3}^{(n+1)}, ...$ by applying equation (\ref{eq:shrinkage01}) to $i=1,2, ...$ in order. The cell contraction process is only performed on cells with porosity exceeding $\epsilon_\mathrm{max}$. When water vapor condenses into ice, the porosity decreases, and when calculated using equation (\ref{eq:shrinkage02}), it becomes $\chi_i > 1$, although in this case, it is assumed that $\chi_i = 1$ and equation (\ref{eq:shrinkage01}) is calculated. This cell becomes a state where $\epsilon < \epsilon_\mathrm{max}$ (the state where the pore is slightly packed).

The time increment, $\Delta t$, was determined so that the eccentric anomaly angle, $\theta$, would be evenly spaced. This is because it improves the resolution around the perihelion point rather than making the intervals of $\Delta t$ equal. Let the interval of $\theta$ be $\Delta \theta$. If the value of $\theta$ at a given calculation step $n$ is denoted as $\theta^{(n)}$, then the corresponding time $t^{(n)}$ can be obtained from equation (\ref{eq:Kepler01}). The time $t^{(n+1)}$ of the next calculation step can be ascertained by substituting $\theta^{(n+1)} = \theta^{(n)} + \Delta \theta$ into equation (\ref{eq:Kepler01}). From this, the time step can be determined as $\Delta t = t^{(n+1)} - t^{(n)}$. Here, the initial value of $\Delta \theta$ was set to be $\Delta \theta = 2 \pi/3000$. When the radius of the primitive core shrinks to 95\% of the initial radius, $R_0$, the value of $\Delta \theta$ is increased by 0.01\% at each calculation step, up to a maximum of $\Delta \theta = 2 \pi/300$, and then kept constant. The reason for setting $\Delta \theta$ to be small at the beginning of the calculation is that a high time resolution is required to handle the ice sublimation in the outermost cells immediately after the calculation starts. We performed this calculation until the ice in the center of the cometary nucleus was depleted (until $\phi_\mathrm{i,1} < 10^{-3}$), and the time required for this was taken as the numerical solution for the desiccation time, $\tau_\mathrm{des}^\mathrm{(cal)}$.

The physical quantities used in the calculations are summarized in Table \ref{table:input01}. We assumed that all water ice was crystalline, and the initial cometary nucleus temperature was set to 150 K. The initial radius $R_0 = 1000 \ \mathrm{m}$ of the cometary nucleus was divided into $N = 10^3$ cells, so the initial cell width was $\Delta r = 1 \ \mathrm{m}$. The diameter of the ice and rock grains was set to 1 m. This is because if the grain size of the solid component is set to less than the width of the calculation cell, then the discretization accuracy of the vapor pressure distribution will deteriorate significantly \citep{Miura2022ApJ}. For other settings, refer to the references in the table.

\begin{table*}
    \caption{Default values of input parameters used in calculations. The unit in $T$ is K.}
    \begin{center}
        \begin{tabular}{lccc}
            \hline
            Quantity & Notation & Value (unit) & Ref. \\
            \hline \hline
            Number of computational cells & $N$ & $10^3$ & \\
            Initial radius of nucleus & $R_0$ & $1000 \ (\mathrm{m})$ & \\
            Initial temperature of nucleus & $T_0$ & $150 \ (\mathrm{K})$ & \\
            Material density of water ice & $\varrho_\mathrm{i}$ & $1.0 \times 10^3 \ (\mathrm{kg \ m^{-3}})$ & \citet{Miura2022ApJ} \\
            Material density of rock & $\varrho_\mathrm{r}$ & $3.0 \times 10^3 \ (\mathrm{kg \ m^{-3}})$ & \citet{Miura2022ApJ} \\
            Initial volume fraction of ice & $\phi_\mathrm{i0}$ & $0.18$ & \citet{Miura2022ApJ} \\
            Initial volume fractioni of rock & $\phi_\mathrm{r0}$ & $0.02$ & \citet{Miura2022ApJ} \\
            Diameter of ice & $d_\mathrm{i}$ & $1.0 \ (\mathrm{m})$ & \\
            Diameter of rock & $d_\mathrm{r}$ & $1.0 \ (\mathrm{m})$ & \\
            Macroporosity in primitive core & $\epsilon_\mathrm{p}$ & $0.8$ & \citet{Miura2022ApJ} \\
            Macroporosity in dust mantle & $\epsilon_\mathrm{m}$ & $0.6$ & \citet{Miura2022ApJ} \\
            Specific heat of ice & $c_\mathrm{i}$ & $2.1 \times 10^3 \ (\mathrm{J \ kg^{-1} \ K^{-1}})$ & \\
            Specific heat of rock & $c_\mathrm{r}$ & $1.2 \times 10^3 \ (\mathrm{J \ kg^{-1} \ K^{-1}})$ & \citet{CORADINI1997337} \\
            Specific sublimation heat of ice & $H_\mathrm{i}$ & $2.8 \times 10^6 \ (\mathrm{J \ kg^{-1}})$ & \\
            Thermal conductivity of ice & $\kappa_\mathrm{i}$ & $567/T \ (\mathrm{W \ m^{-1} \ K^{-1}})$ & \citet{CORADINI1997337} \\
            Thermal conductivity of rock & $\kappa_\mathrm{r}$ & $4.2 \ (\mathrm{W \ m^{-1} \ K^{-1}})$ & \citet{CORADINI1997337} \\
            Albedo of cometary nucleus & $A$ & $0.05$ & \citet{CORADINI1997337} \\
            Equilibrium vapor pressure of ice & $P_\mathrm{i}$ & $3.56 \times 10^{12} \exp(- \frac{6141.667}{T}) \ (\mathrm{Pa})$ & \citet{Mekler1990ApJ} \\
            Solar constant & $S_\odot$ & $1.37 \times 10^3 \ (\mathrm{W \ m^{-2}})$ & \\
            \hline
         \end{tabular}
         \label{table:input01}
    \end{center}
\end{table*}

\section{Analytic model}
\label{sec:steady_model}

Assuming that the temperature and vapor pressure distributions inside the cometary nucleus are steady, we obtained the analytical solution for the desiccation time, $\tau_\mathrm{des}^\mathrm{(ana)}$. In the analytical model, the solar flux on the cometary nucleus surface was averaged over the orbital period.

\subsection{Steady solution in dust mantle}
\label{sec:analytic_solution}

\begin{figure}
   \begin{center}
      \includegraphics*[width=\linewidth]{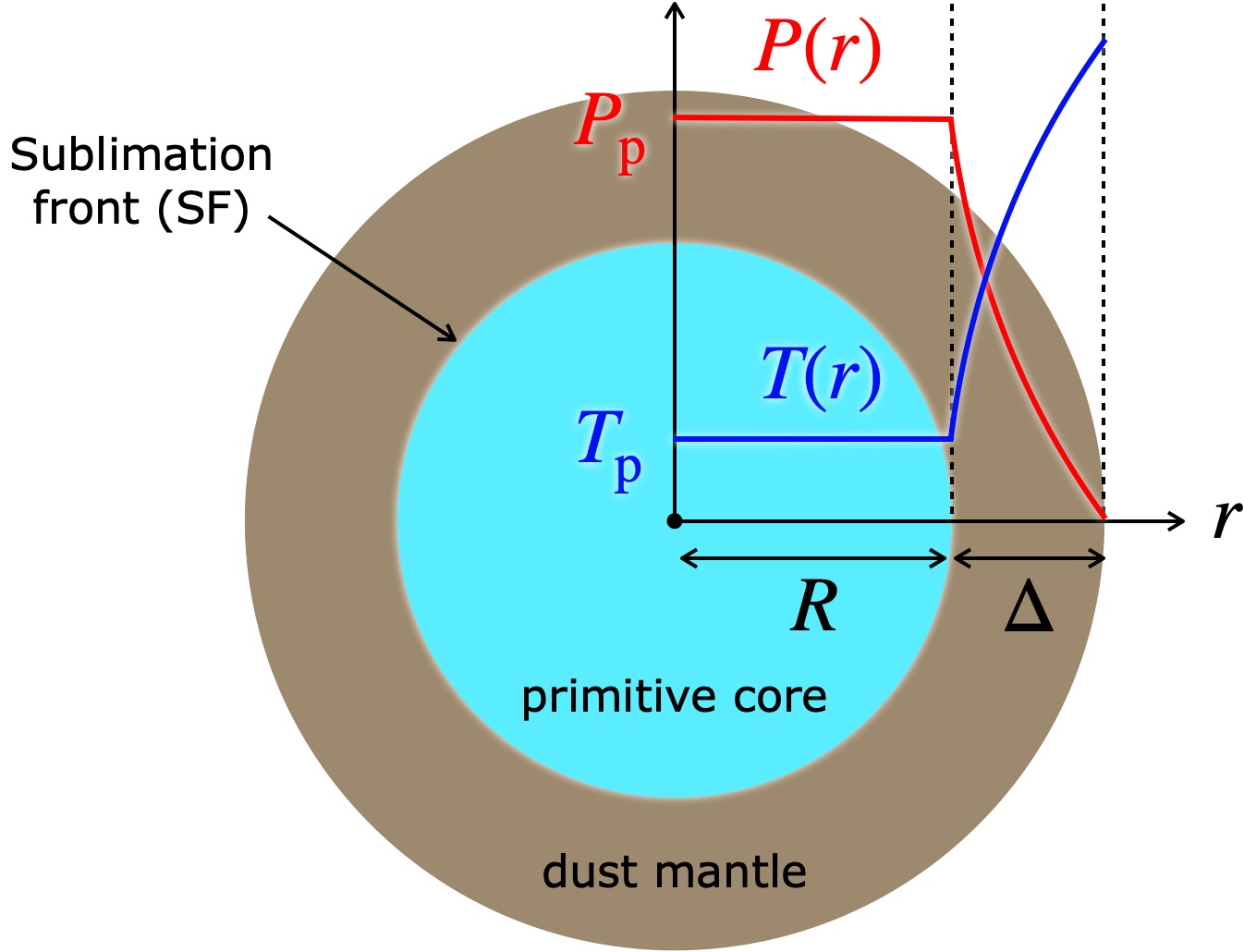}
   \end{center}
   \caption{Schematic diagram of the steady-state distribution of temperature and vapor pressure inside a cometary nucleus. The radius of the primitive core and the thickness of the dust mantle are $R$ and $\Delta$, respectively. The temperature and vapor pressure inside the primitive core are constant at $T_\mathrm{p}$ and $P_\mathrm{p}$, respectively. Ice sublimation only at the surface of the primitive core is considered (sublimation front).
   {Alt text: Illustrations.}}
   \label{fig:distribution_simple}
\end{figure}

Figure \ref{fig:distribution_simple} shows the distributions of temperature and vapor pressure inside the cometary nucleus in a steady state. The interior of the nucleus is assumed to have a two-layer structure: a primitive core and a dust mantle with remaining initial ice and completely depleted ice, respectively. The radius of the primitive core and the thickness of the dust mantle are $R$ and $\Delta$, respectively. The radius of the cometary nucleus is $R_* = R + \Delta$. The solar energy incident from the surface of the dust mantle is transported inward in it by solid heat conduction and is consumed as the sublimation heat of ice on the surface of the primitive core. No sublimation or condensation of ice occurs in the dust mantle ($q_\mathrm{v} = 0$). From equations (\ref{eq:energy_conservation01}) and (\ref{eq:vapor_conservation01}), $T(r)$ and $P(r)$ in the dust mantle satisfy the following spherically-symmetric Laplace equations:
\begin{equation}
    \frac{\partial}{\partial r} \left( r^2 \frac{\partial T}{\partial r} \right) = 0
    , \quad
    \frac{\partial}{\partial r} \left[ r^2 \frac{\partial}{\partial r} \left( \frac{P}{\sqrt{T}} \right) \right] = 0 .
    \label{eq:steady01}
\end{equation}
Here, we assume that the values of $\kappa$, $\epsilon$, and $\bar{d}_\text{s}$ in the dust mantle are uniform.

The temperature inside the primitive core is assumed to be uniform at $T_\mathrm{p}$. In the steady state, the equilibrium between ice and water vapor is established inside the primitive core, and hence, the water vapor pressure is equal to $P_\mathrm{i}(T_\mathrm{p})$. Therefore, the boundary conditions on the surface of the primitive core are given by
\begin{equation}
    T = T_\mathrm{p},
    \quad
    P = P_\mathrm{i} (T_\mathrm{p}) .
    \qquad (\text{at} ~ r = R)
    \label{eq:boundary01}
\end{equation}
The boundary conditions on the dust mantle surface are given by
\begin{equation}
    \kappa \frac{dT}{dr} = \sigma_\mathrm{SB} (\bar{T}_\mathrm{eq}^4-T^4) 
    , \quad
    P = 0 .
    \qquad (\text{at} ~ r = R + \Delta)
    \label{eq:boundary02}
\end{equation}
Here, $\bar{T}_\mathrm{eq}$ is the orbital mean surface temperature, and it is given by averaging equation (\ref{eq:T_eq01}) over the orbital period as follows \citep{KLINGER1981320}:
\begin{equation}
    \bar{T}_\mathrm{eq}^4 
    \equiv \frac{1}{\tau} \int_0^\tau T_\mathrm{eq}^4 dt
    = \frac{(1-A) S_\odot}{4 \sigma_\mathrm{SB} a^2 \sqrt{1-e^2}} .
    \label{eq:Teq_mean02}
\end{equation}
Furthermore, the energy balance equation is established due to the balance between the heat flux on the surface of the primitive core and the sublimation heat of the ice as follows:
\begin{equation}
    \kappa \frac{dT}{dr} = H_\mathrm{i} J_\mathrm{v} .
    \qquad (\text{at} ~ r = R)
    \label{eq:boundary03}
\end{equation}

Solving equation (\ref{eq:steady01}) gives the following general solutions:
\begin{equation}
    T(r) = C_2 - \frac{C_1}{r},
    \quad
    \frac{P(r)}{\sqrt{T(r)}} = C_4 - \frac{C_3}{r} .
\end{equation}
Here, the integral constants $C_1, C_2, C_3, C_4$ and the primitive core temperature $T_\mathrm{p}$ satisfy the following from the boundary conditions (\ref{eq:boundary01}), (\ref{eq:boundary02}), and  (\ref{eq:boundary03}):
\begin{eqnarray}
    T_\mathrm{p} &=& C_2 - \frac{C_1}{R} , \label{eq:boundary11} \\
    \frac{\kappa}{\sigma_\mathrm{SB}} \frac{C_1}{(R+\Delta)^2}
    &=& \bar{T}_\mathrm{eq}^4 - \left( C_2 - \frac{C_1}{R+\Delta} \right)^4 , \\
    \frac{P_\mathrm{i}(T_\mathrm{p})}{\sqrt{T_\mathrm{p}}} &=& C_4 - \frac{C_3}{R} , \\
    0 &=& C_4 - \frac{C_3}{R+\Delta} , \\
    \kappa C_1 &=& - H_\mathrm{i} D_\mathrm{v} C_3 . \label{eq:boundary15}
\end{eqnarray}

\subsection{Temperature of primitive core}

When $C_1, C_2, C_3, C_4$ are eliminated from equations (\ref{eq:boundary11})--(\ref{eq:boundary15}), the following equation for $T_\mathrm{p}$ is obtained:
\begin{equation}
    \frac{H_\mathrm{i} D_\mathrm{v}}{\sigma_\mathrm{SB} R_0} \frac{P_\mathrm{i}(T_\mathrm{p})}{\sqrt{T_\mathrm{p}}} 
    \frac{x}{(x+k)k}
    = \bar{T}_\mathrm{eq}^4 - \left[ T_\mathrm{p} + \frac{H_\mathrm{i} D_\mathrm{v}}{\kappa} \frac{P_\mathrm{i}(T_\mathrm{p})}{\sqrt{T_\mathrm{p}}} \right]^4 ,
    \label{eq:Tp_ana01}
\end{equation}
where, $x = R/R_0$ and $k = \Delta/R_0$ are the radius of the primitive core and the thickness of the dust mantle respectively normalized by the initial radius of the cometary nucleus. $k$ is given as a function of $x$ in the following \citep{Miura2022ApJ}:
\begin{equation}
    k(f,p,x) = \left[x^3 + \frac{1-f}{p} (1-x^3) \right]^{1/3}-x .
    \label{eq:mantle_thickness01}
\end{equation}
Here, $f = \varrho_\mathrm{i} \phi_\mathrm{i0} / (\varrho_\mathrm{r} \phi_\mathrm{r0} + \varrho_\mathrm{i} \phi_\mathrm{i0})$ is the mass fraction of ice in the solid component of the primitive core , and $p = \varrho_\mathrm{r} \phi_\mathrm{r0} / (\varrho_\mathrm{r} \phi_\mathrm{r0} + \varrho_\mathrm{i} \phi_\mathrm{i0})$ is the density ratio of the dust mantle to the primitive core. Using the values in Table \ref{table:input01}, we obtain $f = 0.75$ and $p=5$. Before ice sublimation, $x=1$ and $k=0$. When the internal ice has completely sublimated, $x=0$ and $k=((1-f)/p)^{1/3}$. From this, it is observed that when the ice is completely depleted from the cometary nucleus, the radius will contract to $((1-f)/p)^{1/3}$ times its initial value.

The fact that $x$ is included in equation (\ref{eq:Tp_ana01}) suggests that $T_\mathrm{p}$ changes as the primitive core shrinks. However, considering the physical properties of a typical cometary nucleus as assumed in this study, it is shown that the left-hand side is negligibly small compared to that of the right-hand side. Substituting the values in Table \ref{table:input01} yields
\begin{equation}
    \left( \frac{H_\mathrm{i} D_\mathrm{v}}{\sigma_\mathrm{SB} R_0} \frac{P_\mathrm{i}(T_\mathrm{p})}{\sqrt{T_\mathrm{p}}} \right)^{1/4}
    \simeq 80 \ \mathrm{K} ,
\end{equation}
where, we used $T_\mathrm{p} = 200 \ \mathrm{K}$. This is sufficiently low for the typical value of $\bar{T}_\mathrm{eq}$ assumed in this study (i.e., $\bar{T}_\mathrm{eq} > 150 \ \mathrm{K}$). Therefore, except for the initial stage where $k$ is very small, the left-hand side can be ignored compared to that of the right-hand side. Then, we can rewrite equation (\ref{eq:Tp_ana01}) as
\begin{equation}
    \bar{T}_\mathrm{eq} 
    = T_\mathrm{p} + \frac{H_\mathrm{i} D_\mathrm{v}}{\kappa} \frac{P_\mathrm{i}(T_\mathrm{p})}{\sqrt{T_\mathrm{p}}} .
    \label{eq:Teq_mean01}
\end{equation}
This is the equation that expresses the relationship between $\bar{T}_\mathrm{eq}$ and $T_\mathrm{p}$. This equation is essentially the same as that in equation (22) in \citet{Yu2019MNRAS}. Since $x$ and $k$ are not included in equation (\ref{eq:Teq_mean01}), $T_\mathrm{p}$ can be considered to be constant during the evolution of the cometary nucleus. \citet{Yu2019MNRAS} assumed that $T_\mathrm{p}$ is constant, although this study is the first to provide a rationale for this assumption.

\begin{figure}
   \begin{center}
      \includegraphics*[width=\linewidth]{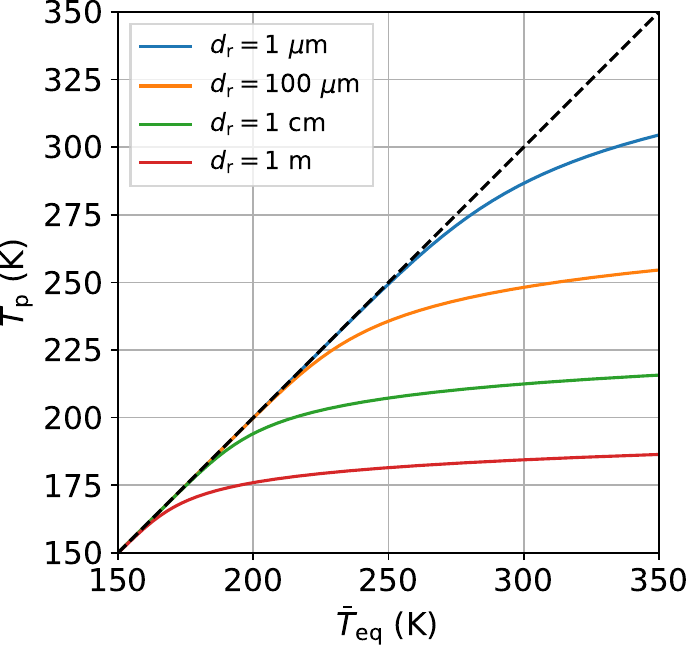}
   \end{center}
   \caption{Relationship between the orbital mean surface temperature $\bar{T}_\mathrm{eq}$ and the primitive core temperature $T_\mathrm{p}$. The different colors of lines represent different diameters of the rocks that comprise the dust mantle, $d_\mathrm{r}$. The values in Table \ref{table:input01} were used for the parameters other than $d_\mathrm{r}$.
   {Alt text: A line graph. The x axis shows the orbital mean surface temperature from 150 to 350 Kelvin. The y axis shows the primitive core temperature with the same range.}}
   \label{fig:Teq02}
\end{figure}

Although $T_\mathrm{p}$ cannot be expressed analytically as a function of $\bar{T}_\mathrm{eq}$, it can be obtained by numerically solving equation (\ref{eq:Teq_mean01}) for a given value of $\bar{T}_\mathrm{eq}$. Figure \ref{fig:Teq02} shows the relationship between $\bar{T}_\mathrm{eq}$ and $T_\mathrm{p}$ for various rock diameters $d_\mathrm{r}$. When $\bar{T}_\mathrm{eq}$ is small, $T_\mathrm{p}$ is almost equal to $\bar{T}_\mathrm{eq}$ because the sublimation heat of ice can be ignored. As $\bar{T}_\mathrm{eq}$ increases, $T_\mathrm{p}$ also increases, although it gradually deviates from the value of $\bar{T}_\mathrm{eq}$. This reflects the effect of the increase in solar flux being offset by the intence sublimation heat of ice. Additionally, for the same value of $\bar{T}_\mathrm{eq}$, the smaller the value of $d_\mathrm{r}$, the larger the value of $T_\mathrm{p}$. This reflects the fact that the smaller the value of $d_{\rm r}$, the more effectively the dust mantle suppresses the outflow of water vapor, and hence, the higher water vapor pressure inside can be maintained.

\subsection{Desiccation time}

The time variation of the primitive core radius is given by the following, using the water vapor flux $J_\mathrm{v}(R)$ on the surface:
\begin{equation}
    \frac{dR}{dt} = - \frac{J_\mathrm{v}(R)}{\varrho_\mathrm{i} \phi_\mathrm{i}} .
    \label{eq:dRdt01}
\end{equation}
$J_\mathrm{v}(R)$ can be derived using the analytical solution of the steady distribution in section \ref{sec:analytic_solution} as follows:
\begin{equation}
    J_\mathrm{v}(R) 
    = - D_\mathrm{v} \left[ \frac{d}{dr} \left( \frac{P}{\sqrt{T}} \right) \right]_{r=R}
    = D_\mathrm{v} \frac{P_\mathrm{i}(T_\mathrm{p})}{\sqrt{T_\mathrm{p}}} \frac{R+\Delta}{R \Delta} .
    \label{eq:vapor_flux_sublimationfront01}
\end{equation}
If we substitute equation (\ref{eq:vapor_flux_sublimationfront01}) into equation (\ref{eq:dRdt01}) and normalize it with $x = R/R_0$ and $k = \Delta/R_0$, we obtain
\begin{equation}
    \frac{dx}{d(t/\tau_\mathrm{sub})}
    = - \left( \frac{1}{k} + \frac{1}{x} \right) ,
    \label{eq:dx/dt01}
\end{equation}
where, $\tau_\mathrm{sub} \equiv (\varrho_\mathrm{i} \phi_\mathrm{i} R_0^2 / D_\mathrm{v}) \sqrt{T_\mathrm{p}} / P_\mathrm{i}(T_\mathrm{p})$. This equation is the same as that in equation derived by \citet{Miura2022ApJ}. Integrating it from 1 to 0 with respect to $x$, the desiccation time is derived as
\begin{equation}
    \tau_\mathrm{des} = \lambda(f,p) \frac{\varrho_\mathrm{i} \phi_\mathrm{i} R_0^2}{D_\mathrm{i}} \frac{\sqrt{T_\mathrm{p}}}{P_\mathrm{i}(T_\mathrm{p})} .
    \label{eq:tau_dep01}
\end{equation}
Here, $\lambda(f,p)$ is a dimensionless quantity reflecting the contraction of the cometary nucleus, and is defined as
\begin{equation}
    \lambda (f,p)
    \equiv \int_0^1 \frac{xk(f,p,x)}{x+k(f,p,x)} dx .
    \label{eq:lambda01}
\end{equation}
The values of $\lambda(f,p)$ calculated by numerical integration are shown in Table \ref{table:lambda}. If the values in Table \ref{table:input01} are substituted, $f = 0.75$ and $p = 5$, we obtain $\lambda = 0.045$.

\begin{table*}
   \caption{Value of $\lambda(f,p)$ obtained by numerically integrating equation (\ref{eq:lambda01}).}
   \label{table:lambda}
   \begin{center}
      \begin{tabular}{|c|ccccccccccc|}
         \hline
         \diagbox{$f$}{$p$} & 1.0 & 1.5 & 2.0 & 2.5 & 3.0 & 3.5 & 4.0 & 4.5 & 5.0 & 5.5 & 6.0 \\
         \hline
         0.05 & 0.164 & 0.142 & 0.127 & 0.117 & 0.108 & 0.101 & 0.096 & 0.091 & 0.087 & 0.083 & 0.080 \\
         0.10 & 0.161 & 0.139 & 0.125 & 0.114 & 0.106 & 0.099 & 0.094 & 0.089 & 0.085 & 0.081 & 0.078 \\
         0.15 & 0.158 & 0.136 & 0.122 & 0.111 & 0.103 & 0.097 & 0.091 & 0.086 & 0.082 & 0.079 & 0.076 \\
         0.20 & 0.154 & 0.133 & 0.119 & 0.109 & 0.101 & 0.094 & 0.089 & 0.084 & 0.080 & 0.077 & 0.074 \\
         0.25 & 0.151 & 0.130 & 0.116 & 0.106 & 0.098 & 0.092 & 0.086 & 0.082 & 0.078 & 0.074 & 0.071 \\
         0.30 & 0.147 & 0.127 & 0.113 & 0.103 & 0.095 & 0.089 & 0.084 & 0.079 & 0.075 & 0.072 & 0.069 \\
         0.35 & 0.143 & 0.123 & 0.109 & 0.100 & 0.092 & 0.086 & 0.081 & 0.076 & 0.073 & 0.070 & 0.067 \\
         0.40 & 0.139 & 0.119 & 0.106 & 0.096 & 0.089 & 0.083 & 0.078 & 0.074 & 0.070 & 0.067 & 0.064 \\
         0.45 & 0.135 & 0.115 & 0.102 & 0.093 & 0.085 & 0.080 & 0.075 & 0.071 & 0.067 & 0.064 & 0.061 \\
         0.50 & 0.130 & 0.111 & 0.098 & 0.089 & 0.082 & 0.076 & 0.071 & 0.068 & 0.064 & 0.061 & 0.059 \\
         0.55 & 0.125 & 0.106 & 0.094 & 0.085 & 0.078 & 0.072 & 0.068 & 0.064 & 0.061 & 0.058 & 0.056 \\
         0.60 & 0.119 & 0.101 & 0.089 & 0.080 & 0.074 & 0.068 & 0.064 & 0.061 & 0.057 & 0.055 & 0.052 \\
         0.65 & 0.113 & 0.095 & 0.084 & 0.075 & 0.069 & 0.064 & 0.060 & 0.057 & 0.054 & 0.051 & 0.049 \\
         0.70 & 0.106 & 0.089 & 0.078 & 0.070 & 0.064 & 0.059 & 0.056 & 0.052 & 0.050 & 0.047 & 0.045 \\
         0.75 & 0.098 & 0.082 & 0.071 & 0.064 & 0.059 & 0.054 & 0.051 & 0.048 & 0.045 & 0.043 & 0.041 \\
         0.80 & 0.089 & 0.074 & 0.064 & 0.057 & 0.052 & 0.048 & 0.045 & 0.042 & 0.040 & 0.038 & 0.036 \\
         0.85 & 0.078 & 0.064 & 0.056 & 0.050 & 0.045 & 0.042 & 0.039 & 0.036 & 0.034 & 0.033 & 0.031 \\
         0.90 & 0.064 & 0.052 & 0.045 & 0.040 & 0.036 & 0.033 & 0.031 & 0.029 & 0.027 & 0.026 & 0.025 \\
         0.95 & 0.045 & 0.036 & 0.031 & 0.027 & 0.025 & 0.023 & 0.021 & 0.020 & 0.018 & 0.017 & 0.017 \\
         \hline
      \end{tabular}
   \end{center}
\end{table*}

\begin{figure}
   \begin{center}
      \includegraphics*[width=\linewidth]{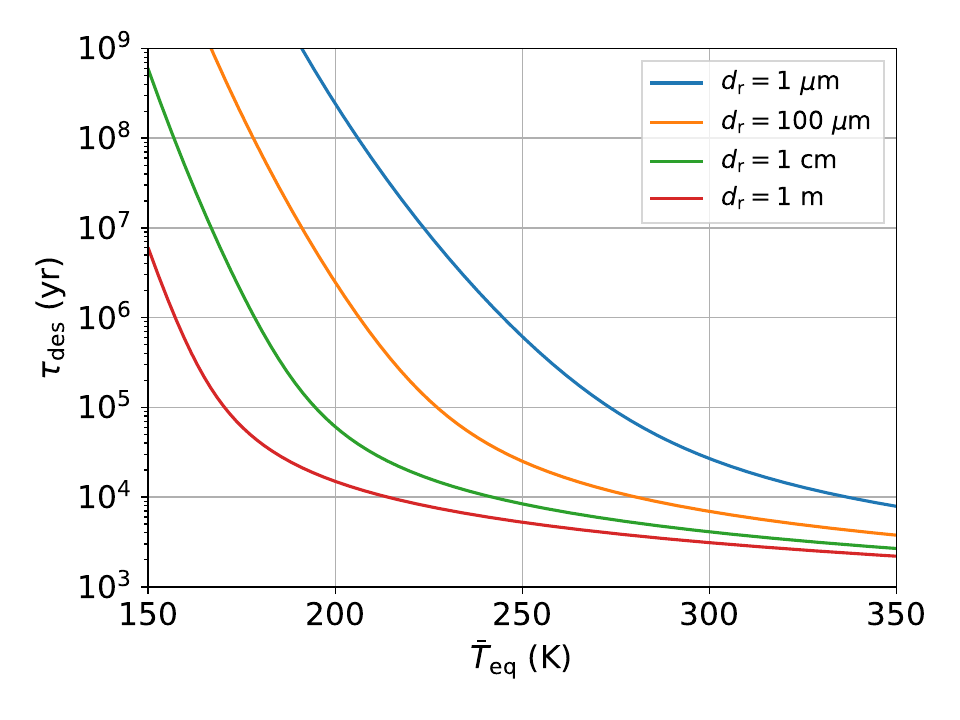}
   \end{center}
   \caption{Analytical solution of desicattion time, $\tau_\mathrm{des}$, as a function of the orbital mean surface temperature, $\bar{T}_\mathrm{eq}$. The meaning of the colors of lines is the same as in Figure \ref{fig:Teq02}.
   {Alt text: A line graph. The x axis shows the orbital mean surface temperature from 150 to 350 Kelvin. The y axis shows the desiccation time from one thousand to one billion years  in logarithmic scale.}}
   \label{fig:tau_dep01}
\end{figure}

Figure \ref{fig:tau_dep01} shows the dependence of $\tau_\mathrm{des}$ on $\bar{T}_\mathrm{eq}$ for various rock diameters $d_\mathrm{r}$. As $\bar{T}_\mathrm{eq}$ increases, the $\tau_\mathrm{des}$ decreases. This implies that the energy required for ice sublimation can be injected in a shorter time due to the increase in the solar flux. Additionally, when $\bar{T}_\mathrm{eq}$ is the same, the smaller the $d_\mathrm{r}$, the larger is the $\tau_\mathrm{des}$. This reflects the fact that when the grain size of the rocks in the dust mantle becomes smaller, it becomes more difficult for water vapor to escape. However, the dependence of $\tau_\mathrm{des}$ on $d_\mathrm{r}$ is weak. For example, when the value of $\bar{T}_\mathrm{eq} = 250$ K, even if the value of $d_\mathrm{r}$ is reduced 100 times from 1 cm to 100 $\mu$m, the desiccation time only increases from $8.2 \times 10^3$ years to $2.5 \times 10^4$ years, a mere 3 times. The weak dependence on $d_\mathrm{r}$ is due to the $d_\mathrm{r}$ dependence of $T_\mathrm{p}$. As $d_\mathrm{r}$ decreases, $T_\mathrm{p}$ increases, increasing the water vapor pressure inside the cometary nucleus, and subseequently the action of discharging water vapor through the dust mantle increases. Therefore, for a cometary nucleus with a radius of about 1 km, if $d_\mathrm{r} \gtrsim 100 \ \mathrm{\mu m}$, then at a temperature of $T_\mathrm{eq} \gtrsim 230 \ \mathrm{K}$, the desiccation time is relatively short, on the order of $10^5$ years or less, and does not depend much on $d_\mathrm{r}$.

\subsection{Water vapor emission rate}

Let $J_\mathrm{v}(R+\Delta)$ be the water vapor flux on the dust mantle surface. If sublimation and condensation of ice in the dust mantle can be ignored, then from the conservation law of the water vapor flux permeating the dust mantle, $R^2 J_\mathrm{v}(R) = (R+\Delta)^2 J_\mathrm{v}(R+\Delta)$ holds. Therefore, we obtain
\begin{equation}
    J_\mathrm{v}(R+\Delta) 
    = \left( \frac{R}{R+\Delta} \right)^2 J_\mathrm{v}(R)
    = D_\mathrm{v} \frac{P_\mathrm{i}(T_\mathrm{p})}{\sqrt{T_\mathrm{p}}} \frac{R}{(R+\Delta)\Delta} .
    \label{eq:vapor_flux_surface01}
\end{equation}
Multiplying the surface area of the cometary nucleus by equation (\ref{eq:vapor_flux_surface01}), we obtain the mass of water vapor $Q_\mathrm{v}$ flowing out from the surface per unit time (vapor emission rate) as follows:
\begin{equation}
    Q_\mathrm{v} = 4 \pi (R + \Delta)^2 J_\mathrm{v}(R+\Delta)
    = 4 \pi D_\mathrm{v} \frac{P_\mathrm{i}(T_\mathrm{p})}{\sqrt{T_\mathrm{p}}} \frac{R(R+\Delta)}{\Delta} .
    \label{eq:vapor_production01}
\end{equation}

\subsection{Maximum size of rocks for ejection}

Due to the ice sublimation, the rocks within the cometary nucleus are subjected to an outward-directed aerodynamic force caused by the vapor flow. If this exceeds the gravity of the cometary nucleus acting on the rock, then the rock will be blown away. The larger the rock, the stronger the gravitational force relative to the aerodynamic force, and hence, an upper limit to the size of rocks that can be blown away exists. Let us denote this diameter $d_\mathrm{max}$. In our model, the vapor flow inside the cometary nucleus also exists, and hence, $d_\mathrm{max}$ can be defined not only on the surface of the cometary nucleus but also inside it. Since the magnitude of aerodynamic and gravitational forces depends on $r$, $d_\mathrm{max}$ also changes depending on $r$. If we ignore the centrifugal force due to the spin of the cometary nucleus, we obtain the following equation from the balance between aerodynamic and gravitational forces as follows \citep{Miura2022ApJ}:
\begin{equation}
    d_\mathrm{max}(r) = \frac{2 v_\mathrm{th}(r)}{\epsilon^{2/3} G \rho_\mathrm{r}}
    \frac{r^2 J_\mathrm{v}(r)}{M(r)} ,
    \label{eq:d_max01}
\end{equation}
where, $v_\mathrm{th}(r) = \sqrt{8 k_\mathrm{B}T(r) / (\pi m_\mathrm{v}})$ is the sound speed of water vapor at position $r$, $J_\mathrm{v}(r)$ is the vapor flux at position $r$, $M(r)$ is the total mass of the cometary nucleus inside position $r$, and $G$ is the gravitational constant. Here, we consider $d_\mathrm{max}$ at the surface of the primitive core ($r = R$) and at the surface of the cometary nucleus ($r = R + \Delta$). $J_\mathrm{v}(R)$ and $J_\mathrm{v}(R+\Delta)$ are given by equations (\ref{eq:vapor_flux_sublimationfront01}) and (\ref{eq:vapor_flux_surface01}), respectively. $M(R) = (4/3) \pi \rho_\mathrm{p} R^3$ and $M(R+\Delta) = M(R) + (4/3) \pi \rho_\mathrm{m} [(R+\Delta)^3-R^3]$. Here, $\rho_\mathrm{p} = \varrho_\mathrm{i} \phi_\mathrm{i0} + \varrho_\mathrm{r} \phi_\mathrm{r0}$ and $\rho_\mathrm{m} = \varrho_\mathrm{r} (1 - \epsilon_\mathrm{m})$ are the densities of the primitive core and the dust mantle, respectively. Substituting these into the equation, we obtain
\begin{eqnarray}
    d_\mathrm{max}(R) &=& \frac{3}{2 \pi}
    \frac{v_\mathrm{th}(R)D_\mathrm{v}}{\epsilon_\mathrm{m}^{2/3} G \varrho_\mathrm{r} \rho_\mathrm{p} R_0^2 }
    \frac{P_\mathrm{i}(T_\mathrm{p})}{\sqrt{T_\mathrm{p}}}
    \frac{x+k}{x^2k} , \\
    d_\mathrm{max}(R+\Delta) &=& \frac{3}{2 \pi}
    \frac{v_\mathrm{th}(R+\Delta)D_\mathrm{v}}{\epsilon_\mathrm{m}^{2/3} G \varrho_\mathrm{r} \rho_\mathrm{p} R_0^2 }
    \frac{P_\mathrm{i}(T_\mathrm{p})}{\sqrt{T_\mathrm{p}}} \nonumber \\
    && \times \frac{(x+k)x}{\left[ x^3 + (1-f)(1-x^3) \right] k} . 
    \label{eq:dmax_MS}
\end{eqnarray}

\subsection{Terminal velocity of ejected rock}

On the surface of the cometary nucleus, rocks with a diameter smaller than that of $d_\mathrm{max}$ given by equation (\ref{eq:dmax_MS}) can be blown away by the escaping water vapor. The rock leaving the surface is accelerated as it moves away from the nucleus, the flux of escaping water vapor also weakens as it moves away, and hence, the rock eventually reaches a certain speed $v_\infty$ (terminal velocity). Subsequently, we calculate $v_\infty$ assuming that the initial velocity when it leaves the surface is zero. 

Let $F_\mathrm{D}$ be the aerodynamic force exerted on the rock by the water vapor flow on the surface of the cometary nucleus, and let $F_\mathrm{G}$ be the gravity. In the following, $v_\mathrm{th}$, $J_\mathrm{v}$, $M$, and $d_\mathrm{max}$ are assumed to be values on the surface of the cometary nucleus (at $r = R + \Delta$). The aerodynamic force acting on the rock with diameter $d_\mathrm{r}$ is given by \citep{Huebner2006,Miura2022ApJ}
\begin{equation}
    F_\mathrm{D} = \frac{\pi}{3}  v_\mathrm{th} d_\mathrm{r}^2 J_\mathrm{v} ,
\end{equation}
where, we set  $\epsilon = 1$ because considering the outer region of the cometary nucleus. The gravitational force is given by
\begin{equation}
    F_\mathrm{G} = \frac{G M m_\mathrm{r}}{(R+\Delta)^2} ,
\end{equation}
where, $m_\mathrm{r} = (\pi/6) d_\mathrm{r}^3 \varrho_\mathrm{r}$ is the mass of the rock. Using $d_\mathrm{max}$ given by equation (\ref{eq:dmax_MS}), the resultant ourward force received by the rock on the surface of the cometary nucleus can be calculated as
\begin{equation}
    F_\mathrm{D}-F_\mathrm{G} = \frac{\pi}{3} v_\mathrm{th} d_\mathrm{max}^2 J_\mathrm{v}
    \left[ \left( \frac{d}{d_\mathrm{max}} \right)^2 - \left( \frac{d}{d_\mathrm{max}} \right)^3 \right] .
\end{equation}

Outside the cometary nucleus, both of these forces decay inversely proportional to $r^2$. Therefore, the resultant force $F(r)$ is rewritten as
\begin{equation}
    F(r) = \left( \frac{R+\Delta}{r} \right)^2 (F_\mathrm{D}-F_\mathrm{G}) .
\end{equation} 
The work $W$ done by this force as it carries the rock from the surface of the cometary nucleus to infinity is
\begin{equation}
    W = \int_{R+\Delta}^{\infty} F(r) dr
    = (R + \Delta) (F_\mathrm{D}-F_\mathrm{G}) .
\end{equation}
As this $W$ becomes the kinetic energy of the rock, the terminal velocity satisfies $W = (1/2) m_\mathrm{r} v_\infty^2$. From this, we obtain
\begin{equation}
    v_\infty = \sqrt{\frac{2W}{m_\mathrm{r}}}
    = \sqrt{ \frac{4 (R+\Delta) v_\mathrm{th} J_\mathrm{v}}{\varrho_\mathrm{r} d_\mathrm{max}} \left( \frac{d_\mathrm{max}}{d_\mathrm{r}} - 1 \right) } .
    \qquad (\text{for} \ d < d_\mathrm{max})
    \label{eq:v_inf01}
\end{equation}
According to this equation, $v_\infty$ approaches infinity in the limit $d_\mathrm{r} \to 0$, although this is physically incorrect. Equation (\ref{eq:v_inf01}) cannot be applied to tiny rocks where $v_\infty$ exceeds the sound speed of water vapor.

\subsection{Elapsed time from initial cometary state to current state}

From the observed value of the current radius $R_*$ of the small celestial body and the assumed value of the dust mantle thickness $\Delta$, the time it takes for the initial state to reach the current state can be calculated if this small body originates the cometary nucleus as below.

Let the ratio of the dust mantle thickness to the radius of the primitive core be defined as $\alpha \equiv k/x = \Delta/R = \Delta/(R_* - \Delta)$. Substituting this into equation (\ref{eq:mantle_thickness01}) and eliminating $k$, we obtain
\begin{equation}
    % x = \left[ \frac{(1-f)/p}{(\alpha+1)^3-1+(1-f)/p} \right]^{1/3} .
    x = \left[ \frac{(1+\alpha)^3-1}{1-f}p + 1 \right]^{-1/3} .
    \label{eq:x_estimate01}
\end{equation}
This allows us to calculate $x$ and the initial radius $R_0 = R_*/x$. Finally, in equation (\ref{eq:dx/dt01}), instead of integrating $x$ from 1 to 0, if we integrate from 1 to $x$, the elapsed time can be calculated from the initial cometary state to the current state in the current orbit as follows:
\begin{equation}
    \tau_\mathrm{elap} = \lambda'(f,p,x) \frac{\varrho_\mathrm{i} \phi_\mathrm{i} R_0^2}{D_\mathrm{i}} \frac{\sqrt{T_\mathrm{p}}}{P_\mathrm{i}(T_\mathrm{p})} ,
    \label{eq:tau_dep02}
\end{equation}
where,
\begin{equation}
    \lambda' (f,p,x)
    \equiv \int_x^1 \frac{xk(f,p,x)}{x+k(f,p,x)} dx .
    \label{eq:lambda02}
\end{equation}

\section{Results and discussions}

\label{sec:results}

\subsection{Time evolution of internal structure of nucleus}

\begin{figure*}
   \begin{center}
      \includegraphics[width=0.45\linewidth]{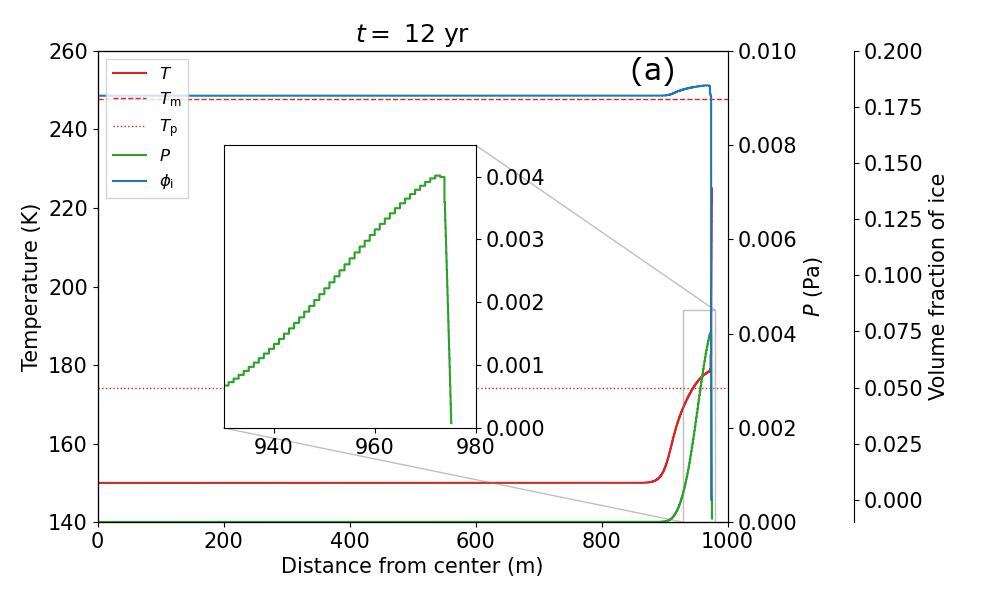}
      \includegraphics[width=0.45\linewidth]{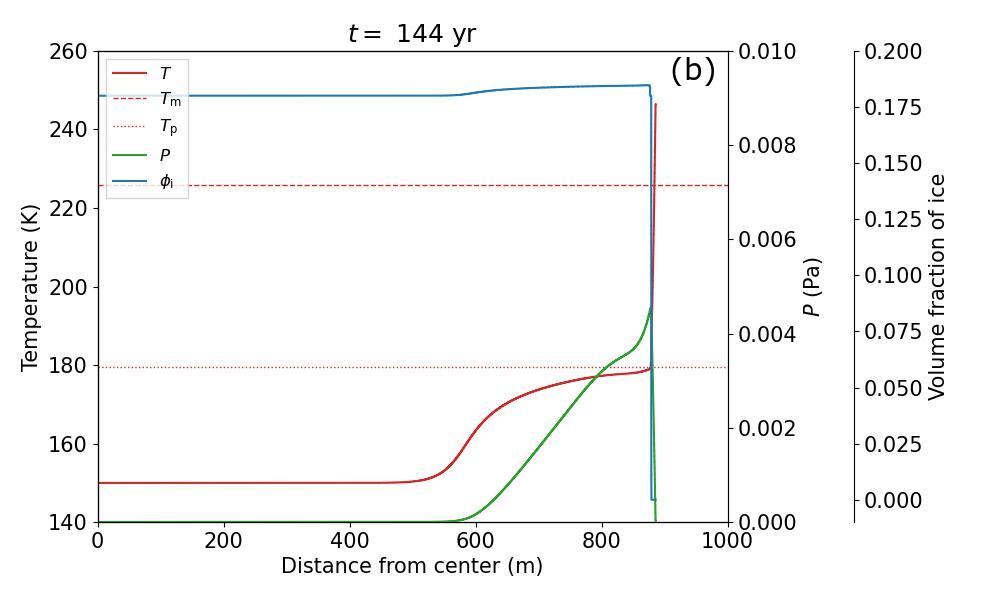}
      \includegraphics[width=0.45\linewidth]{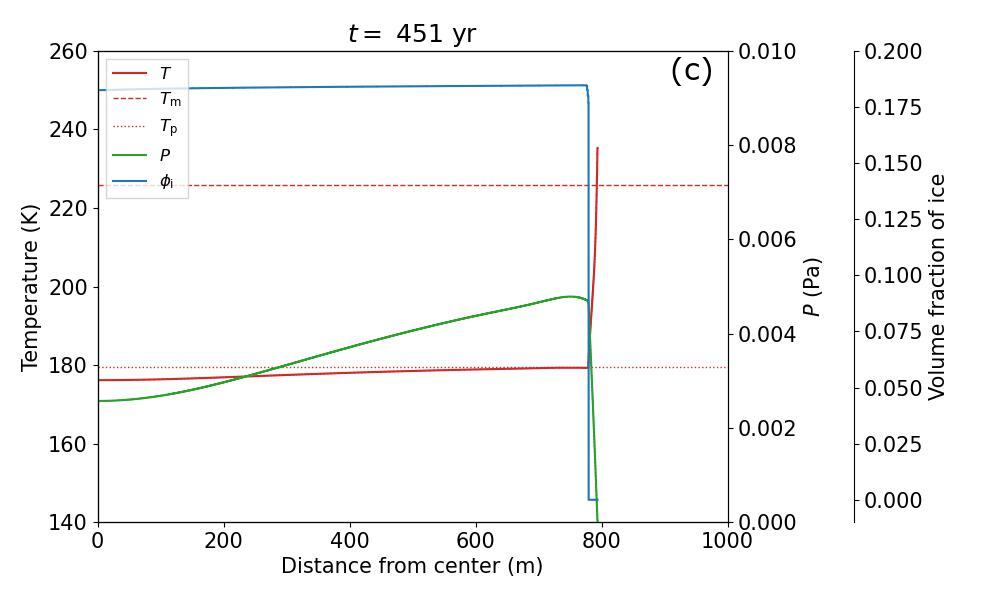}
      \includegraphics[width=0.45\linewidth]{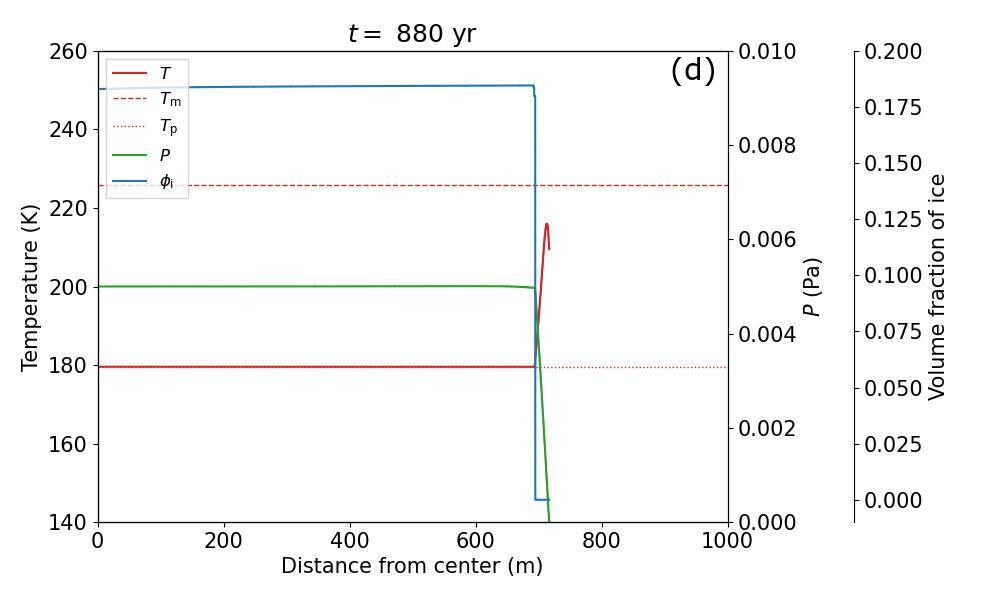}
      \includegraphics[width=0.45\linewidth]{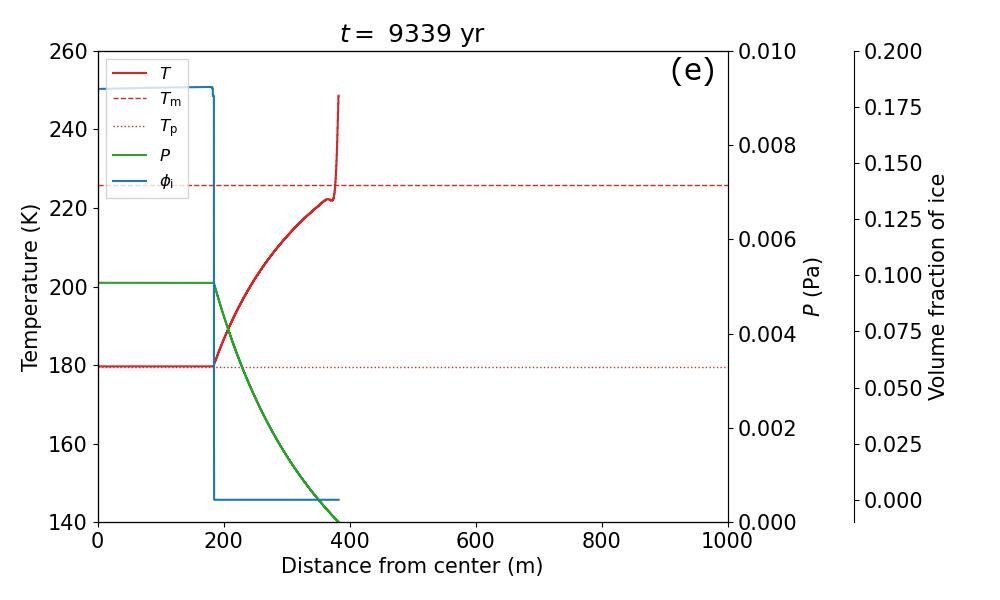}
      \includegraphics[width=0.45\linewidth]{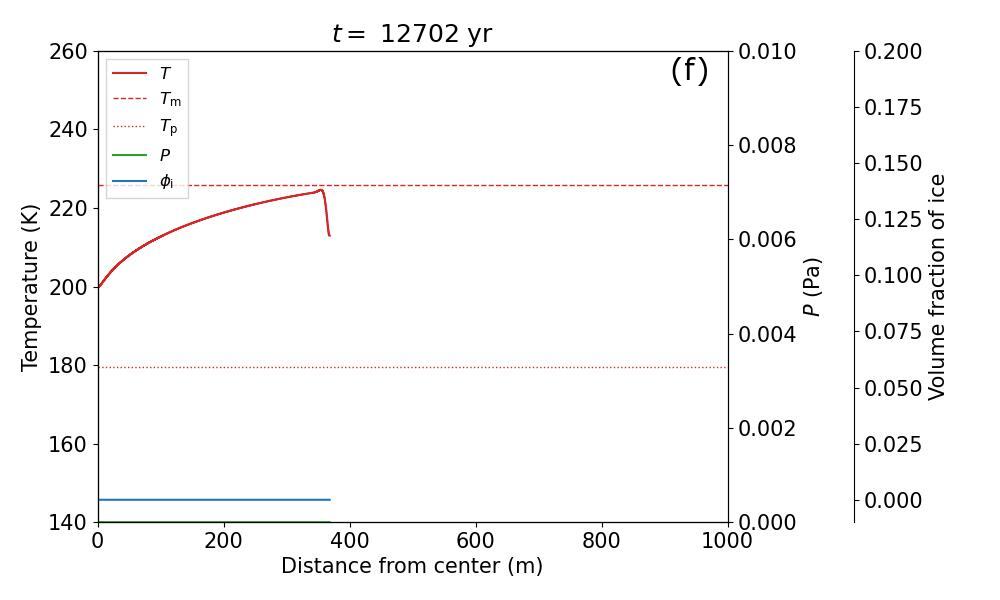}
   \end{center}
   \caption{Numerical results of the time evolution of the internal structure of a cometary nucleus in the case of $a = 1.5$ au and $e = 0.2$. The solid lines represent the temperature $T$ (red), water vapor pressure $P$ (green), and ice volume fraction $\phi_\mathrm{i}$ (blue), respectively. The horizontal axis is the distance $r$ from the center of the nucleus. The dashed line is the orbital mean surface temperature $\bar{T}_\mathrm{eq}$. The dotted line shows the primitive core temperature $T_\mathrm{p}$ calculated based on an analytical model. (a) The nucleus is heated from the surface, then some of the water vapor is transported to the inside of the nucleus. The inset shows an enlarged view of the pressure distribution near the surface. (b) The water vapor transported to the inside condenses, and the temperature increases due to the condensation heat. (c) The temperature rise due to condensation heat reaches the center of the nucleus. (d) The temperature of the primitive region inside the nucleus becomes almost uniform. (e) The temperature of the primitive region remains uniform and constant, and the outflow of water vapor progresses, causing the nucleus to shrink. (f) The ice inside is almost completely depleted.
   {Alt text: Six line graphs. The x axis shows the distance from the nucleus center from 0 to one thousand meters. There are three y axes, the temperature from 140 to 260 Kelvin, the vapor pressure from 0 to 0.01 pascal, and the volume fraction of ice from 0 to 0.2.}}
   \label{fig:result01}
\end{figure*}

Figure \ref{fig:result01} shows the numerical results of the time variation of the internal structure of the cometary nucleus for the case of $a = 1.5$ au and $e = 0.2$. Immediately after the beginning of the calculation, the temperature near the surface increases, and the water vapor pressure increases accordingly (a). Sublimation of ice and contraction of the nucleus proceed, and at this point the radius contracts from 1000 m to approximately 975 m. In the enlarged image, a negative water vapor pressure gradient is generated from the surface to a depth of several meters, and water vapor flows outward. Below that depth, the vapor pressure has a positive gradient, and the water vapor moves towards the inner, colder region, where it is re-condensed into ice. In the region where the water vapor is re-condensed, the temperature rises due to the condensation heat (b). Over time, the region where ice remains (the primitive core) shrinks, and simultaneously, the dust mantle (the region where $\phi_\mathrm{i} \simeq 0$) thickens (c). After approximately 800 years, the temperature of the primitive core becomes almost uniform (d). This is extremely fast compared with that of the time required for heat to be transmitted to the center by solid heat conduction ($R_0^2 / (\kappa/(\varrho c)) \sim 10^5 \ \mathrm{yr}$). The heat transport in the primitive core is conducted by the transport of water vapor and the release and uptake of latent heat associated with sublimation. The cyclic temperature fluctuations near the surface reflect changes in the distance from the Sun, although the primitive core is almost unaffected by the seasonal cycle. When the core temperature is calculated based on the analytical model, we obtain $T_\mathrm{p} = 179.4$ K for the orbital element assumed in this calculation. The core temperature obtained by numerical calculation was in good agreement with the analytical solution (dotted line). The outflow of water vapor continued (e), and the internal ice was almost completely depleted in approximately 12,700 years (f).

\begin{figure}
   \begin{center}
      \includegraphics[width=\linewidth]{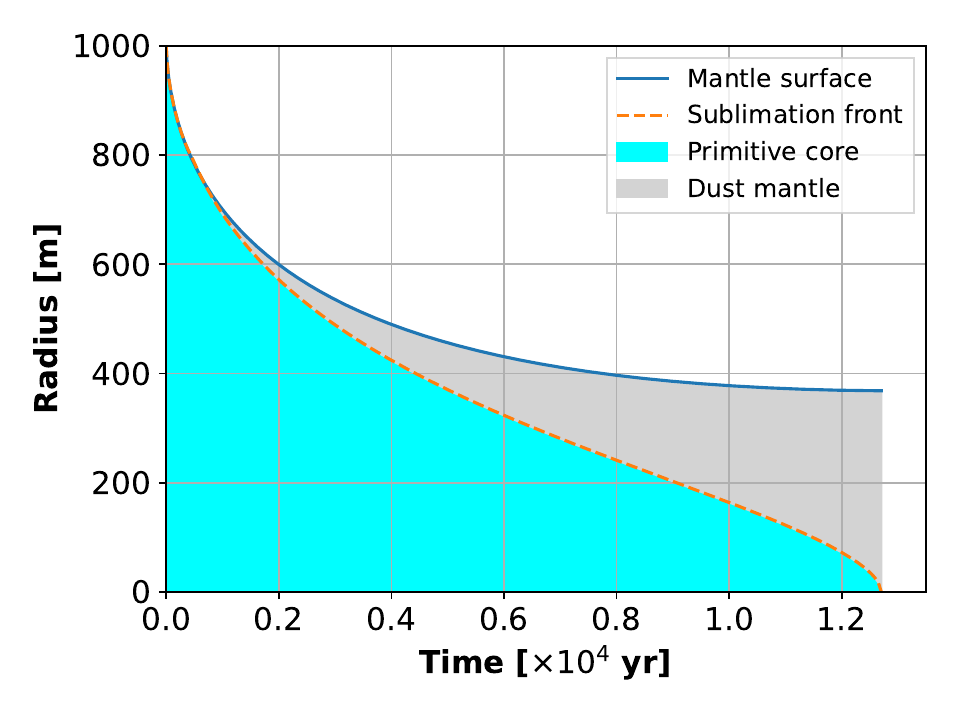}
   \end{center}
   \caption{Numerical results of the contraction of the cometary nucleus due to water ice sublimation. The calculation conditions correspond to those in Figure \ref{fig:result01}. The time variations of the radii of the entire nuclues and the primitive core are shown by solid and dashed cures, respectively.
   {Alt text: A line graph. The x axis shows the elapsed time from 0 to 14000 years. The y axis shows the radius from 0 to 1000 meters.}}
   \label{fig:time_evolution01}
\end{figure}

Figure \ref{fig:time_evolution01} shows the time evolution of the radii of the entire cometary nucleus and the primitive core. As the ice sublimates, the cometary nucleus contracts rapidly, and simultaneously the thickness of the dust mantle increases. The contraction of the cometary nucleus slows down over time, although the ice sublimates at a nearly constant rate inside, and the primitive core continues to shrink. The primitive core radius reached zero at $t = 1.27 \times 10^4$ years, which is the numerical solution for the desiccation time under these conditions. Eventually, a small celestial body consisting of only rocks with a radius of just under 400 m was left behind.

\subsection{Maximum rock size for ejection}

\begin{figure}
   \begin{center}
      \includegraphics[width=\linewidth]{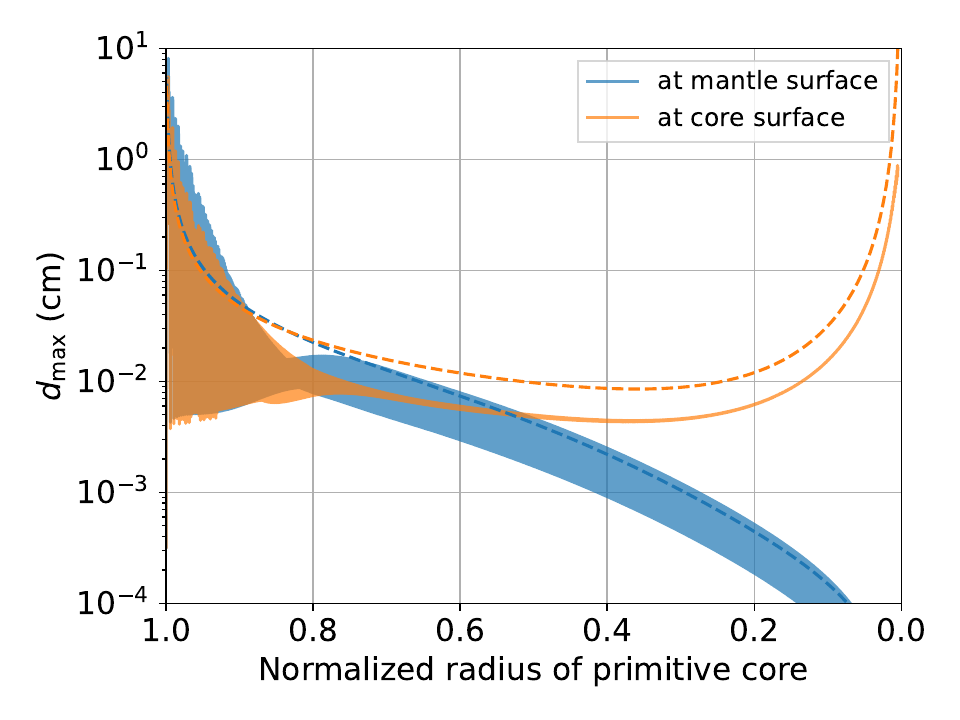}
   \end{center}
   \caption{Numerical results of the time variation of the maximum diameter $d_\mathrm{max}$ of rocks that can be blown away by sublimation gas. The calculation conditions correspond to those in Figure \ref{fig:result01}. The horizontal axis shows the value of the primitive core radius $R$ normalized by the initial radius $R_0$, and evolves from left to right in the figure. The blue line shows the value on the surface of the dust mantle, and the orange line shows the value on the surface of the primitive core. The thickness of the lines indicates the seasonal fluctuation range. The dashed line shows the analytical solution.
   {Alt text: A line graph. The x axis shows the normalized radius of primitive core from 1.0 to 0.0. The y axis shows the maximum diameter of rocks from 1 micrometer to ten centimeters.}}
   \label{fig:main_dmax}
\end{figure}

Figure \ref{fig:main_dmax} shows the time variation of the maximum diameter $d_\mathrm{max}$ of the rock that can be blown away by the outflowing gas. The values at the surface of the cometary nucleus (blue line) and the surface of the primitive core (orange line) are plotted, respectively. If we focus at the surface of the cometary nucleus, then the value of $d_\mathrm{max}$ is the largest at the beginning of the calculation, reaching approximately 10 cm. Subsequently, it decreases gradually. This reflects the fact that the flux of outflowing gas decreases as the dust mantle thickens. Additionally, throughout the entire evolution period, $d_\mathrm{max}$ never exceeds the diameter of the rock $d_\mathrm{r} = 1 \ \mathrm{m}$ assumed in this calculation. This result confirms that the assumption in our numerical model that the rocks are not blown away is valid under this condition. Contrarily, if we focus at the surface of the primitive core, initially $d_\mathrm{max}$ decreases with time in the same way as at the surface of the cometary nucleus, although a subsequent increase is observed. This reflects the fact that the surface of the primitive core penetrates deep into the interior of the cometary nucleus, and the gravity acting on the rocks at that location becomes smaller. A large $d_\mathrm{max}$ on the surface of the primitive core implies that rocks in size not blown off at the surface of the cometary nucleus can be at the surface of the primitive core. As the dust mantle is layered on top of the primitive core, it is not necessarily the case that rocks blown off the surface of the primitive core are directly ejected into space. However, rocks smaller than the size of the pores in the dust mantle may be able to pass through the pores along with the outflowing gas. This suggests the possibility that dust is ejected not only from the surface of the nucleus, but also from inside it.

\subsection{Orbital dependence of desiccation time}

We investigated the dependence of the desiccation time on orbital elements. Numerical calculations were performed for each orbital element by changing the semi-major axis in the range $a = 1.0$--3.0 au in steps of 0.5 au and the eccentricity in the range $e = 0.0$--0.8 in steps of 0.2, until the ice in the center of the cometary nucleus was depleted, and obtained the numerical solution for the desiccation time, $\tau_\mathrm{des}^\mathrm{(cal)}$ . The results of the calculations are summarized in Table \ref{table:sublimation_time}. Additionally, the analytical solution for the desiccation time, $\tau_\mathrm{des}^\mathrm{(ana)}$, is also included in the table. The numerical is systematically larger than the analytical solution, although it is in good agreement with it under many conditions. However, the difference between the numerical and analytical solutions increased as $a$ and $e$ increased, and in some cases the numerical solution was more than several times larger than that of the analytical. Particularly, in the case with the largest $a$ and $e$ ($a = 3.0$ au and $e = 0.8$) considered in this study, the numerical solution was $1.47 \times 10^6$ years, which is about 26 times larger than that of the analytical solution of $5.67 \times 10^4$ years. These results show that the analytical model can predict the desiccation time of the cometary nucleus with high accuracy, although limitations exist in the range of application due to orbital elements.

\begin{table*}
   \caption{Dependence of the desiccation time, $\tau_\mathrm{des}$, on orbital elements. $T_\mathrm{p}$ is the primitive core temperature calculated based on an analytical model, and $\tau_\mathrm{des}^\mathrm{(ana)}$ is the analytical solution for the desicattion time. $\tau_\mathrm{des}^\mathrm{(cal)}$ is the numerical solution for the desiccation time. The relative error between the numerical and analytical solutions is evaluated using $(\tau_\mathrm{des}^\mathrm{(cal)}-\tau_\mathrm{des}^\mathrm{(ana)})/\tau_\mathrm{des}^\mathrm{(ana)}$.}
   \begin{center}
      \begin{tabular}{ccccccccc}
         \hline
         \# & $a$ (au) & $e$ & $\tau$ (yr) & $\bar{T}_\mathrm{eq}$ (K) & $T_\mathrm{p}$ (K) & $\tau_\mathrm{des}^\mathrm{(ana)}$ (yr) & $\tau_\mathrm{des}^\mathrm{(cal)}$ (yr) & error \\
         \hline \hline
         A10E08 & 1.0 & 0.8 & 1.00 & 312.7 & 185.0 & $4.13 \times 10^3$ & $6.95 \times 10^3$ & 0.68 \\
         A10E06 & 1.0 & 0.6 & 1.00 & 291.0 & 184.0 & $4.93 \times 10^3$ & $6.63 \times 10^3$ & 0.34 \\
         A10E04 & 1.0 & 0.4 & 1.00 & 281.3 & 183.5 & $5.39 \times 10^3$ & $6.48 \times 10^3$ & 0.20 \\
         A10E02 & 1.0 & 0.2 & 1.00 & 276.6 & 183.2 & $5.64 \times 10^3$ & $6.40 \times 10^3$ & 0.13 \\
         A10E00 & 1.0 & 0.0 & 1.00 & 275.2 & 183.1 & $5.73 \times 10^3$ & $6.38 \times 10^3$ & 0.11 \\
         \hline
         A15E08 & 1.5 & 0.8 & 1.84 & 255.3 & 181.9 & $7.18 \times 10^3$ & $1.41 \times 10^4$ & 0.96 \\
         A15E06 & 1.5 & 0.6 & 1.84 & 237.6 & 180.5 & $9.24 \times 10^3$ & $1.33 \times 10^4$ & 0.44 \\
         A15E04 & 1.5 & 0.4 & 1.84 & 229.7 & 179.8 & $1.06 \times 10^4$ & $1.29 \times 10^4$ & 0.22 \\
         A15E02 & 1.5 & 0.2 & 1.84 & 225.9 & 179.4 & $1.14 \times 10^4$ & $1.27 \times 10^4$ & 0.11 \\
         A15E00 & 1.5 & 0.0 & 1.84 & 224.7 & 179.3 & $1.16 \times 10^4$ & $1.27 \times 10^4$ & 0.09 \\
         \hline
         A20E08 & 2.0 & 0.8 & 2.83 & 221.1 & 178.9 & $1.25 \times 10^4$ & $3.37 \times 10^4$ & 1.70 \\
         A20E06 & 2.0 & 0.6 & 2.83 & 205.8 & 176.9 & $1.83 \times 10^4$ & $3.06 \times 10^4$ & 0.67 \\
         A20E04 & 2.0 & 0.4 & 2.83 & 198.9 & 175.8 & $2.28 \times 10^4$ & $2.94 \times 10^4$ & 0.29 \\
         A20E02 & 2.0 & 0.2 & 2.83 & 195.6 & 175.2 & $2.58 \times 10^4$ & $2.90 \times 10^4$ & 0.12 \\
         A20E00 & 2.0 & 0.0 & 2.83 & 194.6 & 175.0 & $2.68 \times 10^4$ & $2.88 \times 10^4$ & 0.07 \\
         \hline
         A25E08 & 2.5 & 0.8 & 3.95 & 197.8 & 175.6 & $2.38 \times 10^4$ & $1.34 \times 10^5$ & 4.63 \\
         A25E06 & 2.5 & 0.6 & 3.95 & 184.1 & 172.4 & $4.51 \times 10^4$ & $1.12 \times 10^5$ & 1.48 \\
         A25E04 & 2.5 & 0.4 & 3.95 & 177.9 & 170.3 & $6.93 \times 10^4$ & $1.05 \times 10^5$ & 0.52 \\
         A25E02 & 2.5 & 0.2 & 3.95 & 175.0 & 169.1 & $8.96 \times 10^4$ & $1.04 \times 10^5$ & 0.16 \\
         A25E00 & 2.5 & 0.0 & 3.95 & 174.1 & 168.7 & $9.76 \times 10^4$ & $1.04 \times 10^5$ & 0.07 \\
         \hline
         A30E08 & 3.0 & 0.8 & 5.20 & 180.6 & 171.3 & $5.67 \times 10^4$ & $1.47 \times 10^6$ & 24.9 \\
         A30E06 & 3.0 & 0.6 & 5.20 & 168.0 & 165.4 & $2.00 \times 10^5$ & $1.11 \times 10^6$ & 4.55 \\
         A30E04 & 3.0 & 0.4 & 5.20 & 162.4 & 161.3 & $4.99 \times 10^5$ & $1.04 \times 10^6$ & 1.08 \\
         A30E02 & 3.0 & 0.2 & 5.20 & 159.7 & 159.1 & $8.49 \times 10^5$ & $1.04 \times 10^6$ & 0.22 \\
         A30E00 & 3.0 & 0.0 & 5.20 & 158.9 & 158.4 & $1.01 \times 10^6$ & $1.05 \times 10^6$ & 0.04 \\
         \hline
      \end{tabular}
   \end{center}
   \label{table:sublimation_time}
\end{table*}

\begin{figure*}
   \begin{center}
      \includegraphics[width=0.45\linewidth]{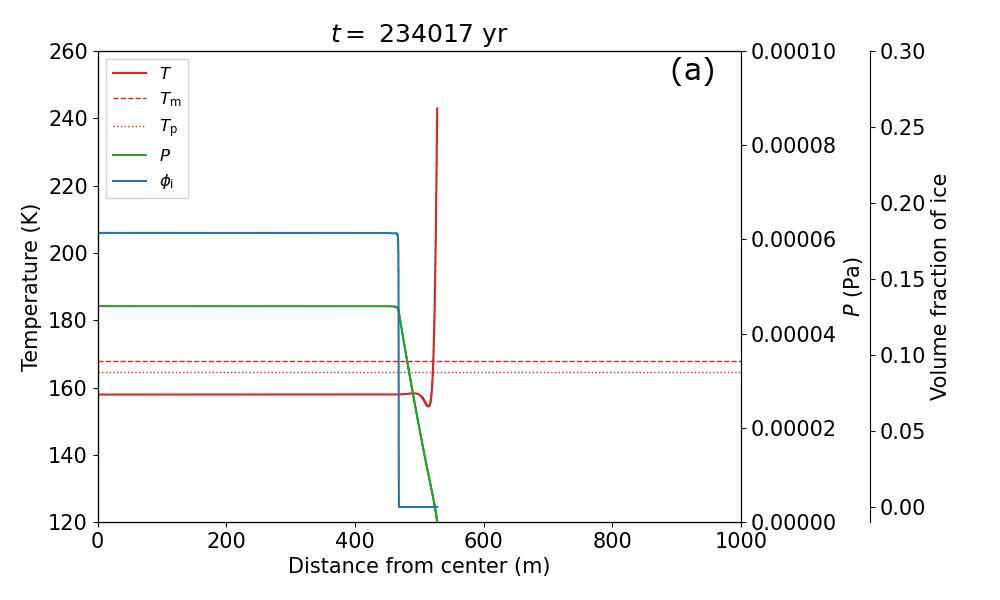}
      \includegraphics[width=0.45\linewidth]{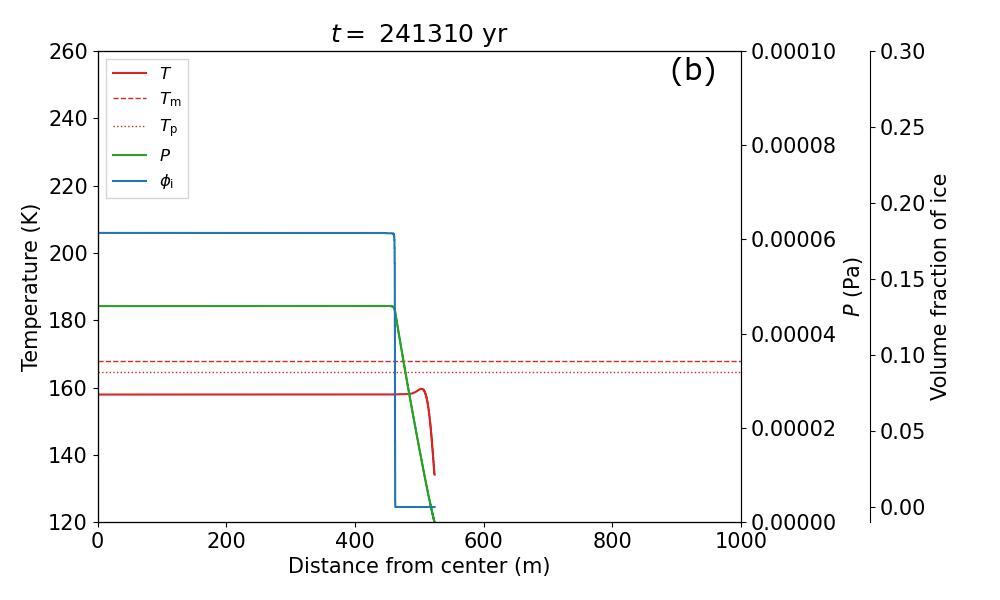}
   \end{center}
   \caption{Temperature distribution inside the cometary nucleus in the case of $a = 3.0$ au and $e = 0.6$: (a) Distribution at perihelion, (b) distribution at aphelion. The meaning of each line is the same as that in Figure \ref{fig:result01}.
   {Alt text: Two line graphs. x and y axes are the same with figure \ref{fig:result01}.}}
   \label{fig:result03}
\end{figure*}

To investigate the cause of the difference between the numerical and analytical solutions for the desiccation time, we compared the temperature distribution inside the cometary nucleus at perihelion and at aphelion when it is in a highly eccentric orbit. Figure \ref{fig:result03} shows the temperature distribution at perihelion (a) and at aphelion (b) for the case of $a = 3.0$ au and $e = 0.6$. It was observed that the difference in the surface temperature is extremely large between perihelion and aphelion. The primitive core temperature is uniform, as in Figure \ref{fig:result01}, although it is considerably lower than that of the analytical model. Because the primitive core temperature is low, it is thought that ice sublimation is suppressed, and it takes longer for ice to be depleted. This result suggests that the analytical model may be inapplicable in cases where the orbital eccentricity is large.

\subsection{Failure of the analytic model}
\label{sec:failure_steady_model}

Here, we consider the reason for the inapplicability of analytical model when the orbital eccentricity is large. When the eccentricity is large, the nuclues recieves most of the sunlight energy during perihelion passage. However, due to the insulating effect of the dust mantle, it takes time for this heat to reach the primitive core. Simultaneously, the nucleus moves away from the Sun. As the nucleus spends a considerable amount of time in the cold region away from the Sun, if the surface temperature of the nucleus is lower than that of the core temperature at that time, it is thought that some of the solar energy received will escape into space before it reaches the primitive core.

Let us find the conditions under which the surface temperature of the cometary nucleus at the aphelion, $T_\mathrm{eq,ap}$, becomes lower than that of the temperature of the primitive core, $T_\mathrm{p}$. Substituting $\theta = \pi$ into equations (\ref{eq:T_eq01}) and (\ref{eq:heliocentric01}), we obtain
\begin{equation}
    T_\mathrm{eq,ap} = \frac{(1-A) S_\odot}{4 \sigma_\mathrm{SB} a^2 (1+e)^2} .
\end{equation}
Contrarily, $T_\mathrm{p}$ can be calculated as a function of $a$ and $e$ based on the analytical model (see equation \ref{eq:Teq_mean01}). Therefore, we can show the region on the $a$--$e$ plane where $T_\mathrm{eq,ap} < T_\mathrm{p}$.

\begin{figure}
   \begin{center}
      \includegraphics[width=\linewidth]{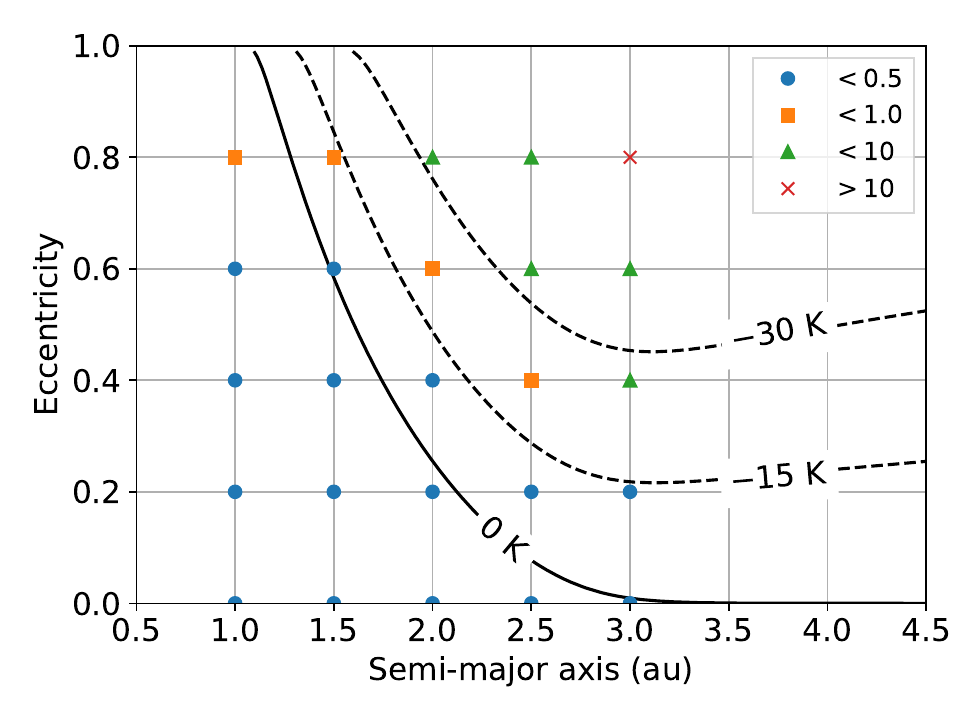}
   \end{center}
   \caption{Condition of orbital elements in which the surface temperature of the cometary nucleus at the aphelion, $T_\mathrm{eq,ap}$, is lower than that of the primitive core temperature, $T_\mathrm{p}$. The black line shows the contour lines of the temperature difference $T_\mathrm{eq,ap} - T_\mathrm{p}$: 0 K, $-15$ K, and $-30$ K. The different relative error between the numerical and analytical solutions is shown using different markers: blue circles, orange squares, green triangles, and red crosses, respectively, for values of error less than 0.5, less than 1.0, less than 10, and greater than 10. See Table \ref{table:sublimation_time} for the values of error.
   {Alt text: A line graph. The x axis shows the semi-major axis from 0.5 to 4.5 astronomical units. The y axis shows the eccentricity from 0 to 1.}}
   \label{fig:e_crit01}
\end{figure}

Figure \ref{fig:e_crit01} shows the orbital elements for which $T_\mathrm{eq,ap}$ is lower than that of $T_\mathrm{p}$. We calculated $T_\mathrm{eq,ap}$ and $T_\mathrm{p}$ for various values of $a$ and $e$, and then plotted the contours of $T_\mathrm{eq,ap} - T_\mathrm{p}$ on the $a$--$e$ plane. As $a$ and $e$ increase, the surface temperature when the nucleus is far from the Sun becomes lower than that of the primitive core, with which the analytical model will be inapplicable. As a rough guide, if $T_\mathrm{eq,ap}$ is about 30 K lower than that of $T_\mathrm{p}$, it is safe to consider that the actual desiccation time will be significantly longer than that of the analytical solution. Conversely, if $T_\mathrm{eq,ap} > T_\mathrm{p}$, the analytical model can predict the desiccation time with an error of approximately a factor of 2.

\subsection{Vapor production rate}

\begin{figure*}
   \begin{center}
      \includegraphics[height=0.4\linewidth]{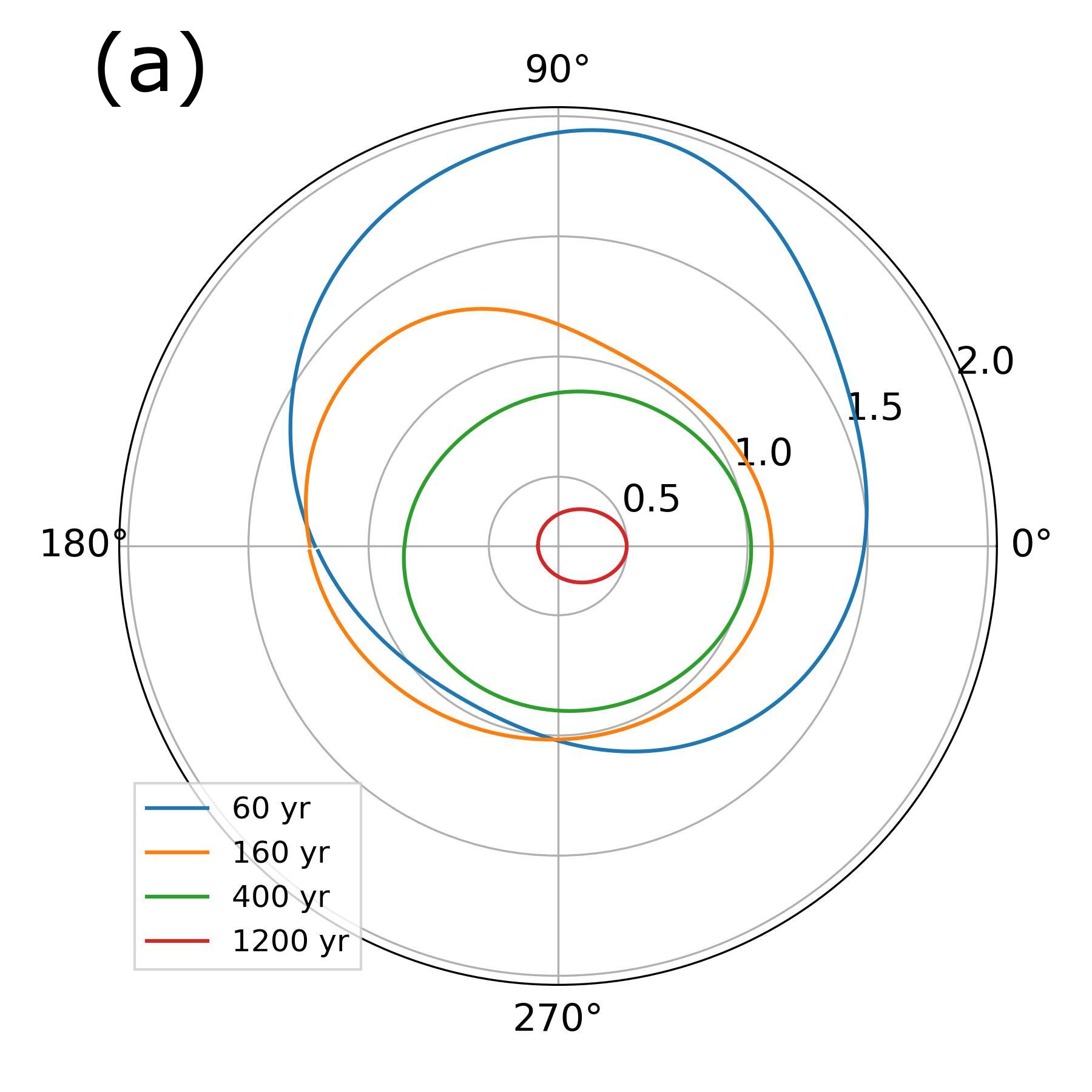}
      \includegraphics[height=0.4\linewidth]{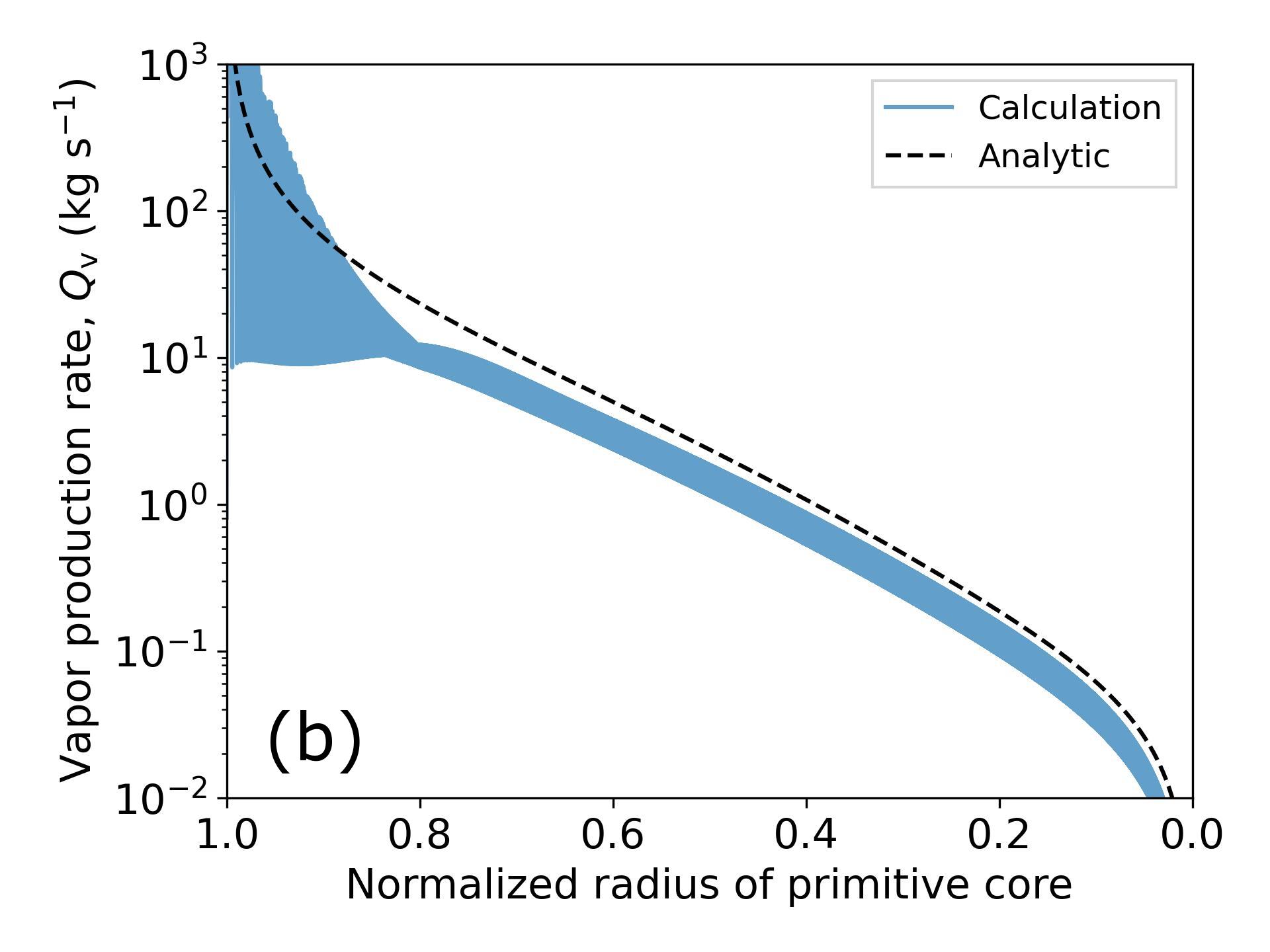}
   \end{center}
   \caption{(a) Dependence of the water vapor emission rate, $Q_\mathrm{v}$, on the eccentric anomaly angle, $\theta$, in the case of $a = 1.5$ au and $e=0.8$. The polar coordinates for $\theta$ are used, with $\theta = 0^\circ$ representing the perihelion and $\theta = 180^\circ$ representing the aphelion. In this figure, the cometary nucleus is orbiting counterclockwise. Each line represents the values of $\log_{10} Q_\mathrm{v} \ \mathrm{(kg \ s^{-1})} = 0.5$, 1.0, 1.5, and 2.0 for one orbital period, at each time. (b) Time variation of $Q_\mathrm{v}$. The horizontal axis shows the primitive core radius $R$ normalized by the initial radius $R_0$, and the time evolution is from left to right in the figure. The blue area shows the numerical results, and the black dashed line shows the analytical solution. The spread of the blue area shows the seasonal variation range.
   {Alt text: Two line graphs. In the right panel, the x axis shows the normalized radius of primitive core from 1.0 to 0.0. The y axis shows the vapor production rate from 0.01 to 1000 kilograms per second.}}
   \label{fig:main_Qv_all}
\end{figure*}

Figure \ref{fig:main_Qv_all} shows the numerical results for the water vapor emission rate $Q_\mathrm{v}$ for the case of $a = 1.5$ au and $e = 0.8$. $Q_\mathrm{v}$ shows seasonal variation in response to changes in the distance from the Sun (a). When $t = 60$ years, $Q_\mathrm{v}$ increases as the cometary nucleus approaches the Sun. Even after the perihelion passage, $Q_\mathrm{v}$ continues to increase, and reaches a peak at around $80$ degrees. The reason for $Q_\mathrm{v}$ peaking after the perihelion passage is that it takes time for the sunlight energy received on the surface to reach the primitive core after being conducted through the dust mantle. This phase lag is further amplified as the dust mantle develops, and at $t = 160$ years, $Q_\mathrm{v}$ is greater at the aphelion than that at the perihelion. At this time, the dust mantle thickness is approximately 6 m, which is about the same as that of the thermal skin depth in the dust mantle ($\sim \sqrt{\kappa / (\rho c)} \simeq 7 \ \mathrm{m}$). When the dust mantle thickness exceeds the thermal skin depth, this phase lag is eliminated, and $Q_\mathrm{v}$ reaches a maximum near the perihelion at $t = 400$ years. At $t = 1200$ years, the mantle thickness reaches approximately 24 m, and the seasonal variation of solar energy is relaxed by the insulation effect of the dust mantle, and hence, the primitive core temperature is almost constant, which implies that the rate of water vapor production on the surface of the primitive core should be almost constant. Nevertheless, the fact that $Q_\mathrm{v}$ changes depending on $\theta$ is due to the seasonal variation in the surface temperature of the dust mantle. At the aphelion, the temperature gradient near the surface of the dust mantle becomes negative. When the temperature gradient is negative, the vapor flux $J_\mathrm{v}$ becomes smaller for the same vapor pressure gradient (see equation {\ref{eq:vapor_flux01}}). This is because the temperature drop causes the vapor to be compressed, and the vapor flow becomes stagnant. Later, as the temperature rises as the nucleus approaches perihelion, the vapor expands, and the outward flux increases. Panel (b) shows the time variation of $Q_\mathrm{v}$. $Q_\mathrm{v}$ monotonically decreases with time while showing seasonal variation. The average trend of the time variation is in good agreement with that of the analytical solution. The reason for the numerical result of $Q_\mathrm{v}$ being systematically smaller than that of the analytical solution is because the temperature of the primitive core is lower than that of the surface temperature at the aphelion, and hence, the solar energy is not efficiently transported to the primitive core (see section \ref{sec:failure_steady_model}).

\section{Implications for individual objects}

\label{sec:discussion}

We applied our analytical model to typical CAT objects and discussed their internal structures and thermal evolutions. The examined CAT objects are summarized in Table \ref{table:application01}. Additionally, we calculated the orbital mean surface temperature, $\bar{T}_\mathrm{eq}$, from the orbital elements of each object, and the primitive core temperature, $T_\mathrm{p}$, based on the analytical model. The grain size in the dust mantle is assumed to be $\bar{d}_\mathrm{s} = 1 \ \mathrm{cm}$, and the values in Table \ref{table:input01} are used for other parameters. For more detailed on these CAT objects, see, for example, \citet{Hsieh2023PSJ}. The following sections consider each CAT object individually.

\begin{table*}
   \caption{Properties of typical CAT objects: semi-major axis $a$, orbital eccentricity $e$, albedo $A$, and current effective radius $R$. The orbital mean surface temperature $\bar{T}_\mathrm{eq}$ calculated from the orbital elements, and the primitive core temperature $T_\mathrm{p}$ calculated based on our analytical model. The grain size of the rocks that make up the dust mantle was set to be $\bar{d}_\mathrm{s} = 1$ cm. The values in Table \ref{table:input01} were used for other physical quantities.}
   \label{table:application01}
   \begin{center}
      \begin{tabular}{lcccccc}
         \hline
         Object & $a$ (au) & $e$ & $A$ & $R$ (m) & $\bar{T}_\mathrm{eq}$ (K) & $T_\mathrm{p}$ (K) \\
         \hline \hline
         (3200) Phaethon & 1.271 & 0.890 & 0.10 & 2500 & 293.1 & 211.9 \\
         107P/(4015) Wilson--Harrington & 2.626 & 0.631 & 0.06 & 1730 & 180.5 & 180.0 \\
         238P/Read & 3.165 & 0.253 & 0.04 & 400 & 156.4 & 156.4 \\
         67P/Churyumov-Gerasimenko & 3.465 & 0.634 & 0.06 & 2000 & 157.3 & 157.3 \\
         133P/Elst-Pizarro & 3.16 & 0.17 & 0.04 & 2500 & 155.8 & 155.8 \\
         \hline
      \end{tabular}            
   \end{center}
\end{table*}

\subsection{(3200) Phaethon}

(3200) Phaethon was originally discovered as an asteroid. Based on its orbital similarity, Phaethon is thought to be the parent body of the Geminids meteor shower \citep{Whipple1983IAUC}. In fact, temporary dust emissions were observed each time it returns to the Sun \citep{Jewitt2010AJ,Jewitt2013ApJ,Hui2017AJ}, although the estimated dust emission rate is an order of magnitude less than that comprising the Geminids meteor shower \citep{Hui2017AJ}. It is thought that the comet activity was once very active in the past, and that the dust released at that time makes up the current Geminids meteor shower \citep{Yu2019MNRAS}. However, the dust ejection mechanism of Phaethon is not well understood. The possibility of ice sublimation was ruled out because the core temperature of Phaethon is estimated to be much higher than that of the sublimation temperature of water ice \citep{Jewitt2006AJ,Jewitt2010AJ}. Contrarily, \citet{Yu2019MNRAS} proposed a long-term thermal evolution model for Phaethon, and concluded that it would take approximately 6 million years for the ice to be completely depleted, and hence, a good chance exists that ice is still inside. Subsequently, we re-examined the possibility that the dust ejection of Phaethon is due to ice sublimation based on our analytical model.

Based on our analytical model, the primitive core temperature expected from orbital elements of Phaethon is $T_\mathrm{p} = 211.9 \ \mathrm{K}$, which is much lower than that of the previous estimates, $T_\mathrm{p} = 300 \ \mathrm{K}$ \citep{Jewitt2006AJ}. This is because previous estimates ignored the sublimation heat of ice. Prior to this study, it was highlighted that the sublimation heat can significantly reduce the core temperature \citep{Schorghofer2018JGR}. Our core temperature estimate is in good agreement with the estimate from the thermal evolution model of \citet{Yu2019MNRAS}.

If we substitute the primitive core temperature into equation (\ref{eq:vapor_production01}), we can calculate the water vapor emission rate $Q_\mathrm{v}$ at the surface of the cometary nucleus. Assuming a dust mantle thickness of 200 m, we obtain $Q_\mathrm{v} = 14 \ \mathrm{kg \ s^{-1}}$. The dust emission rate is likely to be lower than this. Based on observations of comet 238P/Read by the James Webb Space Telescope (JWST) (see the section \ref{sec:Read} for details), the dust-to-gas emission rate ratio is estimated to be approximately 0.3 \citep{Kelley2023Nature}. If we assume that the mass of the ejected dust is 10\% of that of the water vapor, this is in the same order of magnitude as the observationally-based estimates, $\sim 0.1$--1 kg s$^{-1}$ \citep{Jewitt2014AJ,Hui2017AJ}. Additionally, when the primitive core temperature is substituted in equation (\ref{eq:dmax_MS}), the maximum size of dust that can be blown away by ice sublimation, $d_\mathrm{max} = 720 \ \mathrm{\mu m}$, is obtained. The upper limit of dust size emitted from the current Phaethon was estimated to be $65 \ \mathrm{\mu m}$ from observations \citep{Hui2017AJ}, and our estimate can explain the blowing off of dust of this size. Contrarily, based on observations of lunar impact flashes by Geminid meteoroids, it is thought that the Geminid meteor shower contains dust particles of centimeter size \citep{Ortiz2015MNRAS}, although the blowing away of this size cannot be explained by the current water vapor flux.

Let us consider the possibility that comet activity of Phaethon was once more active. Based on equation (\ref{eq:tau_dep02}), we can estimate the period when the comet activity was active. As stated in the previous paragraph, assuming that the current dust mantle thickness is 200 m, the primitive core radius is 2300 m, and we obtain the ratio of the dust mantle thickness to the primitive core radius, $\alpha = 0.0870$. Substituting this in equation (\ref{eq:x_estimate01}) gives $x = 0.531$, and we can estimate the initial radius before ice sublimation to be $R_0 = 4330 \ \mathrm{m}$. Additionally, we obtain $\lambda'(f,p,x) = 8.02 \times 10^{-3}$ by numerically integrating equation (\ref{eq:lambda02}). Substituting this into equation (\ref{eq:tau_dep02}) gives the time required for the dust mantle to develop to its current estimated thickness, $\tau_\mathrm{elap} = 2.17 \times 10^4 \ \mathrm{yr}$. Specifically, if no major changes existed in the orbital elements of Phaethon in past, then it is possible that the activity of Phaethon was extremely high about 20,000 years ago compared to that of the current. Assuming a dust mantle thickness of 1 m at that time, we obtain $Q_\mathrm{v} = 3.1 \times 10^3$ kg s$^{-1}$ and $d_\mathrm{max} = 13$ cm. Compared with that of the current estimate, $Q_\mathrm{v}$ is more than 100 times higher, and it is possible to blow away dust of cm-size contained in the Geminids meteor shower. Furthermore, the estimated ejection timing is consistent with that of the numerical calculations of the formation of the dust stream of the Geminids meteor shower. \citet{Jo2024AA} numerically investigated the dynamics of mm- and cm-size dust ejected from Phaethon at low speeds (on the order of 1 m s$^{-1}$ on Phaethon's Hill sphere), and suggested that dust grains of total mass of $10^{10}$--$10^{14}$ g, corresponding to less than 0.1\% of the mass of Phaethon, were ejected 18,000 years ago. They assumed that the dust was ejected at low speed due to the rotation instability of Phaethon, although sublimation of ice can also produce a similar ejection speed; calculating the terminal velocity of the ejected rock based on equation (\ref{eq:v_inf01}) gives $v_\infty = 0.77 \ \mathrm{m \ s^{-1}}$ for dust with $d = d_\mathrm{max}/2 = 7.5 \ \mathrm{cm}$. These considerations strongly suggest that the dust emission of Phaethon was caused by sublimation of ice.

The DESTINY+ exploration mission, which is aiming to be launched in 2025, plans to perform flyby imaging of Phaethon and in-situ direct analysis of the mass distribution, velocity, direction of arrival, and chemical composition of dust particles in the vicinity of Phaethon \citep{OZAKI202242,TOYOTA2023aa}. In the near future, we will be able to obtain more detailed observational data on the dust emission mechanism from Phaethon. By comparing this with theoretical models, we can expect to gain a better understanding of the formation and evolution of small solar system bodies.

\subsection{238P/Read}
\label{sec:Read}

Comet 238P/Read is the second MBC to be discovered. Since a dust tail was observed at each perihelion passage, it was thought that activity of Read was due to sublimation of ice \citep{Hsieh2006Science,Hsieh2009bAJ}. Water vapor was detected for the first time using the infrared camera of JWST \citep{Kelley2023Nature}. The water vapor emission rate was estimated to be approximately 0.3 kg s$^{-1}$, and the dust emission rate was estimated to be approximately 30\% of that ($\sim 0.1$ kg s$^{-1}$). Additionally, based on observation data from the 2005 perihelion passage, the maximum size of the emitted dust was estimated to be $a_\mathrm{max} =$ (1--10) mm \citep{Hsieh2009bAJ,Hsieh2023PSJ}.

The primitive core temperature of Read is considerably lower than that of Phaethon. To explain the observed water vapor emission rate, it is predicted that the dust mantle was thin. Assuming a mantle thickness of 1 cm, the water vapor emission rate is estimated to be $Q_\mathrm{v} = 0.3$ kg s$^{-1}$ from equation (\ref{eq:vapor_production01}), which is consistent with that of the observation results from JWST. Additionally, from equation (\ref{eq:dmax_MS}), the maximum diameter of the dust that can be blown away is calculated to be $d_\mathrm{max} \simeq 2 \ \mathrm{mm}$, which is in good agreement with the estimated value based on the observation results in 2005. Despite having a smaller $Q_\mathrm{v}$ than that of Phaethon, it is able to blow off larger particles because Read is smaller than Phaethon. The smaller the celestial body, the smaller the surface gravity, while the vapor flux at the surface becomes larger (see equation \ref{eq:vapor_flux_surface01}). Therefore, the smaller the celestial body, the more dust can be blown away with less water vapor. The assumption that the dust mantle is thin makes sense if we consider that it is difficult for a permanent mantle to form because most of the dust is blown away by ice sublimation.

\subsection{133P/Elst-Pizarro}

Comet 133P/Elst-Pizarro (EP) is the first MBC to be discovered. The dust ejection was confirmed near perihelion \citep{Hsieh2004AJ}. The dust emission rate was estimated to be approximately 0.01 kg s$^{-1}$.

Assuming that the current effective radius is equal to the initial radius $R_0$, and substituting the primitive core temperature into equation (\ref{eq:tau_dep01}), the desiccation time of EP is estimated to be approximately 1.2 billion years. This is comparable to the age of the solar system, and hence, it is almost certain that ice exists inside. Assuming a dust mantle thickness of 1 m in equation (\ref{eq:vapor_production01}), the water vapor emission rate is estimated to be approximately $Q_\mathrm{v} \simeq 0.1$ kg s$^{-1}$. If we assume that the dust emission rate is approximately 10\% of that of water vapor, then it is consistent with that based on observation results. Additionally, from equation (\ref{eq:dmax_MS}), the maximum diameter of the dust that can be blown off is estimated to be $d_\mathrm{max} \simeq 3 \ \mathrm{\mu m}$. The dust ejected from EP can be expected to be much smaller than that ejected from Phaethon and Read. \citet{Hsieh2004AJ} estimated that the optically dominant dust particle size ejected from EP was approximately 10 $\mu$m, which is in the same order of magnitude as that of our theoretical prediction.

\subsection{67P/Churyumov-Gerasimenko}

Comet 67P/Churyumov-Gerasimenko (CG) is a typical Jupiter family comet (JFC), and it is known as one of the few comet nuclei that were observed in detail at close range by the ESA comet probe Rosetta. The water vapor emission was confirmed at a distance of 583 million km (3.89 au) from the Sun, and the estimated emission rate was $Q_\mathrm{v} \simeq$ 0.3--1.2 kg s$^{-1}$ \citep{Gulkis2015Science}.

If a dust mantle of a certain thickness is formed, the temperature of the primitive core will be almost constant regardless of the distance from the Sun, and hence, even at the aphelion, a certain amount of water vapor should be produced inside the cometary nucleus. From equation (\ref{eq:vapor_production01}), if we assume that the dust mantle thickness is 1 m, the vapor emission rate is $Q_\mathrm{v} \sim 0.1$ kg s$^{-1}$. This is in agreement with that of the lower limit of the observational results to the order of magnitude. This result suggests that, unless the pores in the dust mantle are somehow blocked, water vapor may be released on a regular basis even at a location sufficiently far from the Sun. One reason for the pores being blocked is the re-condensation of water vapor. When the cometary nucleus moves away from the Sun, the temperature near the surface  of the dust mantle decreases significantly. In the case of the orbit of CG, the surface temperature at the aphelion is more than 30 K lower than that of the primitive core temperature estimated from our analytical model (see Figure \ref{fig:e_crit01}). This suggests that water vapor transported from the primitive core may re-condense near the surface of the dust mantle and fill the pores. In fact, it was observed that ice regenerates in the shaded areas on the surface of CG \citep{De-Sanctis2015Nature}. It is possible that the water vapor that is constantly emitted from the primitive core condenses near the surface of the dust mantle at the aphelion and forms a lid, and as the comet approaches the Sun, the lid slightly sublimates and the water vapor trapped inside begins to be released.

\subsection{107P/Wilson-Harrington}

Comet 107P/Wilson-Harrington (WH) was the first object to be observed in an active comet state and then in an inactive asteroid state \citep{Marsden1992,FERNANDEZ1997114}. Orbital evolution calculations suggested that WH is most likely not of JFC origin, but rather came from the outer main belt \citep{BOTTKE2002399}. Observations estimated its size, albedo, spin period, and shape \citep{Licandro2009AA,URAKAWA201117}. A reanalysis of observation data in 1949 estimated that the ejected material was dust with a size of $44 \ \mathrm{\mu m}$--1.5 mm \citep{Jin2024AA}.

The fact that the comet activity cannot be confirmed regularly each time the perihelion is passed makes it difficult to believe that the activity observed in 1949 was due to ice sublimation. It is likely that no ice is left inside the WH, or if it exists, it is only in a small area near the center. Assuming a dust mantle thickness of 1500 m in equation (\ref{eq:vapor_production01}), the water vapor emission rate is estimated to be $Q_\mathrm{v} \simeq 8 \times 10^{-4} \ \mathrm{kg \ s^{-1}}$. This is more than two orders of magnitude smaller than that of the observed and estimated values for other candidate CATs such as Phaethon, Read, EP, and CG. Additionally, from equation (\ref{eq:dmax_MS}), the maximum diameter of the dust that can be blown off is $d_\mathrm{max} \simeq 0.3 \ \mathrm{\mu m}$. Specifically, it is difficult to blow away dust particles larger than a few tens of micrometers in size due to ice sublimation. The above is consistent with that of the conventional view that WH is a comet that has almost completely exhausted its resources, or that it was originally an asteroid.

\section{Summary}

\label{sec:conclusion}

We proposed a new theoretical model for the thermal evolution of porous cometary nuclei, considering the contraction of the nucleus due to sublimation of volatile substances caused by solar heating. Using an analytical model assuming a seasonally averaged solar heating rate and a steady distribution inside the nucleus, we calculated the time required for the depletion of water ice from the nucleus (desiccation time), the water vapor emission rate, the maximum size of rocks that can be blown away, and the terminal velocity of the blown-away rocks can be evaluated analytically as a function of the orbital elements of the cometary nucleus and the thickness of the dust mantle and so forth. Additionally, the conditions for applying the analytical model were clarified using a numerical model that takes into account the seasonal variation in solar heating rates. Furthermore, the analytical model was applied to typical CAT objects, and the current internal structure and evolutionary processes were considered. The conclusions obtained in this study are as follows.

\begin{enumerate}
    \item The analytical model can accurately predict the desiccation time for cometary nuclei with small orbital semi-major axes and orbital eccentricities. Contrarily, it underestimates the desiccation time for cometary nuclei with large semi-major axes or eccentricities. The conditions under which the analytical model is applicable can be roughly explained by ``the surface temperature of the cometary nucleus at the aphelion being higher than that of the primitive core temperature predicted based on the analytical model.''
    \item When the dust mantle becomes thicker than the thermal skin depth, the temperature of the primitive core inside can be considered to be almost constant and uniform. Even while the cometary nucleus is at its aphelion, a certain amount of water vapor must be being generated inside it. The seasonal variation in the water vapor emission rate is thought to be due in part to the seasonal variation in the temperature near the surface of the nucleus, which causes the water vapor constantly being generated and transported from the inside to expand and contract, and to re-condense and re-sublimate.
    \item The dust emission from the asteroid Phaethon is caused by sublimation of ice. Assuming that the dust mantle thickness is 200 m, the current dust emission rate and the size of the emitted dust, which are estimated based on an analytical model, are in good agreement with that of the observation results. Additionally, if the orbital elements Phaethon have not changed significantly in past, it is estimated that it would take about 20,000 years for the dust mantle to develop to this thickness from an initially primitive cometary state. Specifically, it is thought that Phaethon was significantly more active about 20,000 years ago compared to that of its current state, and that the dust released at that time formed the dust stream of the Geminids meteor shower.
    \item We estimated the current dust mantle thickness for each of the main belt comets Read and Elst-Pizarro, and reproduced the observed results for the water vapor emission rate, dust emission rate, and size of the emitted dust.
    \item The observational evidence that 67P/Churyumov-Gerasimenko was emitting water vapor at a considerable distance from the Sun suggests that this may be due to the steady production of water vapor in the primitive core, where the temperature does not fluctuate seasonally. On the surface of this comet, a cycle of water vapor re-condensation and re-evaporation linked to seasonal fluctuations in the surface temperature may exist.
    \item The comet activity of 107P/Wilson-Harrington, which was observed in 1949, cannot be explained by the ice sublimation.
\end{enumerate}

\begin{ack}
We would like to thank Editage (www.editage.jp) for English language editing.
\end{ack}

\section*{Funding}
H. M. was partly supported by the Grand-in-Aid for Scientific Research [20K05347, 22K18306].

% \section*{Data availability} 
%  The data underlying this article are available ...  

\end{document}